\documentclass[a4paper,12pt]{article}
\usepackage{epsfig}\parskip 5pt plus 1pt
\usepackage{amsmath}
\usepackage{amssymb}
\usepackage{amsfonts}
\usepackage{fleqn}
\usepackage{mathrsfs}
\usepackage{cite}

\def\be{\begin{equation}}
\def\ee{\end{equation}}
\def\bea{\begin{eqnarray}}
\def\eea{\end{eqnarray}}

\begin{document}

\bibliographystyle{OurBibTeX}

\begin{titlepage}

 \vspace*{-15mm}
\begin{flushright}
\today\\
\end{flushright}
\vspace*{5mm}

\begin{center}
{ \bf \Large Minimal $E_6$ Supersymmetric Standard Model}
\\[8mm]
R.~Howl\footnote{E-mail: \texttt{rhowl@phys.soton.ac.uk}.} and
S.F.~King\footnote{E-mail: \texttt{sfk@hep.phys.soton.ac.uk}.}
\\
{\small\it
School of Physics and Astronomy, University of Southampton,\\
Southampton, SO17 1BJ, U.K.\\[2mm]
}
\end{center}
\vspace*{0.75cm}

\begin{abstract}
\noindent
\end{abstract}
We propose a Minimal $E_6$ Supersymmetric Standard Model
(ME$_6$SSM) which
allows Planck scale unification, provides a solution to the
$\mu$ problem and predicts a new $Z'$.
Above the conventional GUT scale
$M_{GUT}\sim 10^{16}$ GeV the gauge group corresponds
to a left-right symmetric Supersymmetric Pati-Salam model,
together with an additional $U(1)_{\psi}$ gauge group
arising from an $E_6$ gauge group broken near the Planck scale.
Below $M_{GUT}$ the ME$_6$SSM contains
three reducible $\mathbf{27}$ representations of the
Standard Model gauge group together with an additional $U(1)_X$ gauge group,
consisting of a novel and non-trivial
linear combination of $U(1)_{\psi}$ and two Pati-Salam
generators, which is broken at the TeV scale by the same singlet
which also generates the effective \(\mu\) term, resulting in a
new low energy \(Z'\) gauge boson.  We discuss the phenomenology of the new \(Z'\) gauge boson in some detail.
\end{titlepage}
\newpage

\section{Introduction}

The question of unification of all the forces of Nature is one of the
most bold and intriguing in all of physics. It would appear to
be somewhat esoteric or premature but for the possibility that
that discoveries at the CERN
LHC may provide an unprecedented opportunity to shed light on this question.
For example, it is well known that, without new physics,
the electroweak and strong gauge couplings
extracted from LEP data and extrapolated to high energies using the
renormalisation group (RG) evolution do not meet within the
Standard Model (SM),
so unification appears to require some new physics.
One example of new physics that can lead to unification is
TeV scale supersymmetry (SUSY). For example, in the
minimal supersymmetric standard model (MSSM) \cite{1},
the gauge couplings converge to a common value at a high energy scale
$M_{GUT}\sim 10^{16}$ GeV, allowing some supersymmetric Grand Unified Theory (GUT)
to emerge above this scale. Thus evidence for the MSSM at the LHC
would provide support for unification at $M_{GUT}$.

However, despite its obvious attractions,
the standard paradigm of SUSY GUTs based on the MSSM faces some
serious shortcomings. On the one hand, the failure to discover
superpartners or the Higgs boson by the LEP and the Tevatron
indicates that the scale of SUSY breaking must be higher
than previously thought, leading to fine-tuning at the per cent level.
On the other hand experimental limits on proton decay
and the requirement of Higgs doublet-triplet splitting
provides some theoretical challenges at the high scale
\cite{Raby:2004br,Raby,Dimopoulos:1981dw}. Related to the doublet-triplet splitting
problem is the origin of $\mu$, the SUSY Higgs
and Higgsino mass parameter, which from phenomenology
must be of order the SUSY breaking scale,
but which {\it a priori} is independent of the SUSY breaking scale.
Apart from these problems, unification of gauge
couplings near $M_{GUT}\sim 10^{16}$ GeV
leaves open the question of a full unification of all the forces
with gravity, although this may be achieved in the framework of
string unification, including high energy threshold effects
\cite{Ross:2004mi}.

The challenges facing SUSY GUTs based on the MSSM motivate
alternative approaches that successfully overcome these problems.
Recently an $E_6$ Supersymmetric Standard Model
(E$_6$SSM) has been proposed \cite{King:2005my,E$_6$SSM}, in which
the low energy particle content consists of three
irreducible $\mathbf{27}$
representations of the gauge group $E_6$ plus, in addition, a pair of
non-Higgs doublets $H',\overline{H}'$ arising from incomplete
$\mathbf{27}',\overline{\mathbf{27}}'$ representations. In the E$_6$SSM
gauge coupling unification works very well, even better than in the MSSM
\cite{King:2007uj}. Moreover, the E$_6$SSM also solves
the $\mu$ problem via a singlet coupling to two Higgs doublets.
In the E$_6$SSM a special role is played by
a low energy gauged $U(1)_N$ symmetry, which consists
of a particular linear combination of Abelian generators contained
in the $SU(5)$ breaking chain of $E_6$. The $U(1)_N$ results
from GUT scale Higgs which develop vacuum expectation values (VEVs)
in their right-handed neutrino components. As a consequence
the right-handed neutrino carries zero charge under $U(1)_N$, thereby allowing
a conventional see-saw mechanism via heavy right-handed Majorana
masses. The $U(1)_N$ is anomaly-free since the
low energy theory contains complete $\mathbf{27}'s$ (minus the heavy
chargeless right-handed neutrinos) and \(H',\overline{H}'\), which have opposite $U(1)_N$ charges.
The $U(1)_N$ also serves to
forbid the $\mu$ term, but allow the singlet coupling, and is broken
by the singlet VEV which generates the effective $\mu$ term.
Without the gauged $U(1)_N$ symmetry there would be a Peccei-Quinn global
symmetry of the theory resulting in an unwanted Goldstone boson.
In the next-to-MSSM (NMSSM) this global symmetry is broken
explicitly by a cubic singlet term, however this only reduces
the symmetry to $Z_3$ and so the singlet VEV then leads to dangerous
cosmological domain walls. In the E$_6$SSM, with a gauged
$U(1)_N$ symmetry, the cubic singlet term is forbidden
and the resulting would-be Goldstone boson
of the theory is eaten by the Higgs mechanism to produce a
massive $Z'$ gauge boson.

However, although the E$_6$SSM solves the $\mu$ problem, the presence of the
non-Higgs doublets $H',\overline{H}'$ cannot be regarded as completely
satisfactory since it introduces a new $\mu'$
problem and in this case a singlet coupling generating the
mass $\mu'$ for $H',\overline{H}'$ is
not readily achieved \cite{King:2005my,E$_6$SSM}.
Similarly, although the E$_6$SSM solves the usual doublet-triplet
splitting problem, since the usual Higgs doublets are contained
along with colour triplets in complete low energy
$\mathbf{27}$ representations,
the presence of the low energy $H',\overline{H}'$ introduces a new
doublet-triplet splitting problem because their triplet partners
are assumed to be very heavy. However, since the
only purpose of including the non-Higgs states $H',\overline{H}'$ is
to help achieve gauge coupling unification at $M_{GUT}\sim 10^{16}$
GeV, it is possible to consider not introducing
$H',\overline{H}$ at all, thereby allowing a solution to
the $\mu$ problem and the doublet-triplet splitting problem,
without re-introducing new ones.
An immediate objection to removing the non-Higgs states
$H',\overline{H}'$ of the E$_6$SSM is that the gauge
couplings will no longer converge at $M_{GUT}\sim 10^{16}$ GeV,
so at first sight this possibility looks
unpromising. In a recent paper we showed how this objection
could be overcome by embedding the theory
into a left-right symmetric
Pati-Salam theory at $M_{GUT}\sim 10^{16}$ GeV,
leading to a unification of all forces with
gravity close to the Planck scale \cite{Howl:2007hq}.
However, the analysis did not include the effects of an
additional low energy $U(1)'$ gauge group that would be required
for a consistent resolution of the $\mu$ problem.

The purpose of the present paper is to
propose a Minimal $E_6$ Supersymmetric Standard Model (ME$_6$SSM)
based on three low energy reducible $\mathbf{27}$ representations of the
Standard Model gauge group which
allows Planck scale unification and provides a solution to the
$\mu$ problem and doublet-triplet splitting problem,
without re-introducing either of these problems.
Above the conventional GUT scale the ME$_6$SSM is
embedded into a left-right symmetric Supersymmetric Pati-Salam model
with an additional $U(1)_{\psi}$ gauge group
arising from an $E_6$ gauge group broken near the Planck scale.
In our previous analysis \cite{Howl:2007hq} we assumed for simplicity that
the $U(1)_{\psi}$ gauge group was broken at the Planck scale.
Here we assume that $U(1)_{\psi}$ remains unbroken
down to $M_{GUT}$ and that below $M_{GUT}$
an additional $U(1)_X$ gauge group,
consisting of a novel and non-trivial
linear combination of $U(1)_{\psi}$ and two Pati-Salam
generators, survives down to low energies. Eventually $U(1)_X$
is broken at the TeV scale by the same singlet
that also generates the effective \(\mu\) term, resulting in a
new low energy \(Z'\) gauge boson.  We discuss the phenomenology of the new \(Z'\) gauge boson in some detail.
We compare the \(Z'\) of the ME$_6$SSM (produced via the
Pati-Salam breaking chain of $E_6$, where $E_6$ is broken
at the Planck scale)
to the  \(Z'\) of the E$_6$SSM (from the $SU(5)$ breaking chain
of $E_6$, where $E_6$ is broken
at the GUT scale) and discuss how they can be distinguished by their
different couplings, which enable the two models to be resolved
experimentally. In the case of the ME$_6$SSM the \(Z'\) gauge boson
properties can be said to provide a window on Planck scale physics.
In both cases the right-handed neutrinos carry
zero charge under the extra low energy $U(1)'$ gauge groups,
allowing a conventional see-saw mechanism.

The layout of the rest of this paper is as follows. In the section 2
we consider the pattern of symmetry breaking assumed in this paper.
In section 3 we consider the two loop RG evolution of gauge
couplings in this model from low energies, through the Pati-Salam
breaking scale at $M_{GUT}\sim 10^{16}$ GeV (assuming various
Pati-Salam breaking Higgs sectors) and show that the Pati-Salam
gauge couplings converge close to the Planck scale $M_p\sim 10^{19}$
GeV. In section 4 we discuss the phenomenology of the new $Z'$
of the ME$_6$SSM and compare it to that of the E$_6$SSM.
In section 5 we shall construct an explicit supersymmetric
model of the kind we are considering.
Finally we conclude the paper in section 5.

\section{Pattern of Symmetry Breaking}

The two step pattern of gauge group symmetry breaking that we analyse in
this paper is:
\begin{equation} \label{eq:steps}
E_6 \overbrace{\longrightarrow}^{M_p} G_{4221} \otimes D_{LR}
\overbrace{\longrightarrow}^{M_{GUT}} G_{3211}
\end{equation}
where the gauge groups are defined by:
\bea
G_{4221} &\equiv & SU(4)
\otimes SU(2)_L \otimes SU(2)_R \otimes U(1)_{\psi}, \nonumber \\
G_{3211} &\equiv &
SU(3)_c \otimes SU(2)_L \otimes U(1)_Y \otimes U(1)_X
\eea
and we have assumed that the
first stage of symmetry breaking happens close to the Planck scale
and the second stage happens close to the conventional GUT scale.
The first stage of symmetry breaking is based on the maximal
\(E_6\) subgroup \(SO(10) \otimes U(1)_{\psi}\) and the maximal
\(SO(10)\) subgroup $G_{422} \otimes D_{LR}$ corresponding to a
Pati-Salam symmetry with \(D_{LR}\) being a discrete left-right
symmetry.\footnote{Under \(D_{LR}\) the matter multiplets transform
as \(q_L \rightarrow q^c_L\), and the gauge groups $SU(2)_L$ and
$SU(2)_R$ become interchanged \cite{Chang:1984uy}.}
The pattern of symmetry breaking assumed in this paper is different
from that commonly assumed in the literature based on the maximal
\(SO(10)\) subgroup $SU(5)\otimes U(1)_{\chi}$
\cite{King:2005my,E$_6$SSM,Hewett:1988xc}. In particular, the
Pati-Salam subgroup does not contain the Abelian gauge group factor
$U(1)_{\chi}$. The only Abelian gauge group factor involved in this
pattern of symmetry breaking is $U(1)_{\psi}$, with a low energy
$U(1)_X$ emerging along with $U(1)_Y$ below the Pati-Salam
symmetry breaking scale.

The first stage of symmetry breaking close to $M_p$ will not be
considered explicitly in this paper. We only remark that the Planck scale
theory may or may not be based on a higher dimensional string
theory. Whatever the quantum gravity theory is, it will involve some
high energy threshold effects, which will depend on the details of
the high energy theory and which we do not consider in our
analysis. Under
\(E_6 \rightarrow G_{4221}\) the fundamental \(E_6\) representation
\(\mathbf{27}\) decomposes as:
\be
\mathbf{27}\rightarrow
(4,2,1)_{\frac{1}{2}} +
(\overline{4}, 1, \overline{2})_{\frac{1}{2}} + (1,2,2)_{-1} +
(6,1,1)_{-1} + (1,1,1)_2
\label{27decomposition}
\ee
where the subscripts are related to the \(U(1)_{\psi}\)
symmetry's charge assignments \cite{Slansky:1981yr,Keith:1997zb}.  One family of
the left-handed quarks and leptons can come from the
\((4,2,1)_{\frac{1}{2}}\); one family of the charge-conjugated quarks
and leptons, including a charge-conjugated neutrino \(\nu^c\), can
come from the \((\overline{4}, 1, \overline{2})_{\frac{1}{2}}\); the
MSSM Higgs bosons can come from the \((1,2,2)_{-1}\); two
colour-triplet weak-singlet particles can come from the
\((6,1,1)_{-1}\); and the \((1,1,1)_{2}\) is a MSSM singlet.

The second stage of symmetry breaking close to $M_{GUT}$ is within
the realm of conventional quantum field theory and requires some
sort of Higgs sector, in addition to the assumed matter content of
three $\mathbf{27}$ representations of the gauge group $E_6$. In
order to break the symmetry \(G_{4221}\) to  \(G_{3211}\) at
$M_{GUT}$, the minimal Higgs sector required are the $G_{4221}$
representations
$\overline{H}_R=(\overline{4},1,\overline{2})_{\frac{1}{2}}$ and
\(H_R=(4,1,2)_{-\frac{1}{2}}\).\footnote{We show in the Appendix that the symmetry breaking \(G_{4221}\) to \(G_{3211}\) also requires an MSSM singlet \(S\), from a \(\mathbf{27}\) multiplet of \(E_6\), to get a low-energy VEV.  The VEV of this MSSM singlet is also used to solve the \(\mu\) problem.}  When these
particles obtain VEVs in the right-handed neutrino directions
$<\overline{H}_R>=<\nu^c_H>$ and $<H_R>=<\nu^H_{R}>$ they
break the \(SU(4) \otimes SU(2)_R \otimes U(1)_{\psi}\)
symmetry to \(SU(3)_c \otimes U(1)_Y \otimes U(1)_X\).
Six of the off-diagonal \(SU(4)\) and two of the
off-diagonal \(SU(2)_R\) fields receive masses related to the VEV of the
Higgs bosons. The gauge bosons associated with the
diagonal \(SU(4)\) generator $T^{15}_4$,
the diagonal \(SU(2)_R\) generator
$T^3_R$ and the
$U(1)_{\psi}$ generator \(T_{\psi}\),
are rotated by the Higgs bosons to create one
heavy gauge boson and two massless gauge bosons associated with
\(U(1)_Y\) and \(U(1)_X\). The part of the symmetry breaking
\(G_{4221}\) to  \(G_{3211}\)
involving the diagonal generators is then:
\be
U(1)_{T^{15}_4}\otimes U(1)_{T^3_R}\otimes U(1)_{\psi}\rightarrow
U(1)_Y \otimes U(1)_X,
\label{sb}
\ee
where the charges of the ``right-handed neutrino''
component of the Higgs which gets the VEV are:
\be
\nu^H_{R}=\left(-\frac{1}{2}\sqrt{\frac{3}{2}}, \ \frac{1}{2},\
-\frac{1}{2}\sqrt{\frac{1}{6}}\right)
\ee
under the corresponding correctly $E_6$ normalized diagonal generators:\footnote{Note that we choose to
normalize the \(E_6\) generators \(G^a\) by \(Tr(G^a G^b) = 3
\delta^{a b}\). It then follows that
the Pati-Salam and standard model operators are conventionally
normalized by \(Tr(T^a T^b) = \frac{1}{2} \delta^{ab}\).
The correctly normalized \(E_6\) generator corresponding to
$U(1)_{\psi}$ is $T_{\psi}/\sqrt{6}$ where $T_{\psi}$
corresponds to the charges in Eq.\ref{27decomposition}.}
\be
T^{15}_4=\sqrt{\frac{3}{2}}~
diag(\frac{1}{6},\frac{1}{6},\frac{1}{6},-\frac{1}{2}),\ \
T^3_R=\frac{1}{2}~ diag(1,-1),\ \ T_{\psi}/\sqrt{6}.
\ee
In the Appendix we discuss the symmetry breaking
in Eq.\ref{sb} in detail.  To simplify the discussion we observe that
$T^{15}=\sqrt{\frac{3}{2}}\frac{(B -L)}{2}$ where
\(B\) and \(L\) are the baryon and lepton number
assignments of each Standard Model particle.
The Higgs charges can then be written as
\be
\nu^H_{R}=\left(-\frac{1}{2}, \ \frac{1}{2},\
-\frac{1}{2}\right)
\ee
under the corresponding generators
$T_{B-L} = \frac{B-L}{2}$, $T^3_R$ and $T_{\psi}$.
Then it is clear to see why the hypercharge generator $Y$
is preserved by the
Higgs \(H_R\) and $\overline{H}_R$ since
\be
Y = T^3_R+\frac{(B -L)}{2}
\label{Y}
\ee
takes a zero value for the right-handed neutrino and anti-neutrino
Higgs components which develop VEVs. However this is not the only
Abelian generator that is preserved by this Higgs sector. The Higgs
\(H_R\) and $\overline{H}_R$ VEVs also preserve the combinations of
generators $T_{\psi}-T_{B-L}$ and $T_{\psi}+T^3_R$.

As shown in the Appendix, precisely one additional Abelian
generator orthogonal to $U(1)_Y$ is preserved, namely:

\be
\label{X}
X=(T_{\psi}+T^3_R)-c_{12}^2Y
\ee
where $c_{12}=\cos \theta_{12}$ and
the mixing angle is given by \be \tan
\theta_{12}=\frac{g_{2R}}{g_{B-L}}, \ \ \ \
g_{B-L}=\sqrt{\frac{3}{2}}\ g_4, \ee where the
$E_6$ normalized Pati-Salam coupling
constants ${g_{2R}}$ and ${g_{4}}$ are evaluated at
the \(G_{4221}\) symmetry breaking scale $M_{GUT}$.  Note that this Abelian generator \(X\) depends on the values that the Pati-Salam coupling constants take at a particular energy scale.  It is easy to prove that it is a general rule that, if three massless gauge
fields are mixed, then at least two of the resulting mass eigenstate
 fields must have a charge that
depends on the value of the original gauge coupling constants.  See the Appendix for more discussion on this unusual aspect of \(X\).

The ``GUT'' (in this case $E_6$)
normalized generator is
\be
T_X={\frac{1}{N_X}}X
\ee

where the normalization constant \(N_X\) is given by:

\begin{equation} \label{eq:NX}
N^2_X \equiv 7 - 2c^2_{12} + \frac{5}{3} c^4_{12}
\end{equation}

From Eq.\ref{X}, the corresponding gauge coupling constant $g^0_X$ may be expressed
in terms of the \(SU(4)\), \(SU(2)_R\) and \(U(1)_{\psi}\) gauge coupling
constants \(g_4,g_{2R}\) and \(g_{\psi}\) as:
\begin{equation} \label{eq:alphaX}
\frac{1}{\alpha^0_X} = \frac{1}{\frac{1}{6}\alpha_{\psi}} + \frac{1}{\frac{3}{2} \alpha_4 + \alpha_{2R}}
\end{equation}
where \(\alpha^0_X = \frac{(g^0_X)^2}{4\pi},~\alpha_{2R} =
\frac{g^2_{2R}}{4\pi}\); \(\alpha_4 = \frac{g^2_4}{4\pi}\); and
\(\alpha_{\psi} = \frac{g^2_{\psi}}{4\pi}\).  In terms of the \(E_6\)
normalized generator \(T_X = X / N_X\), the normalized gauge coupling
constant \(g_X = g^0_X N_X\) so that \(\alpha_X = \alpha_X^0 N_X^2\)
giving:
\begin{equation} \label{eq:alphaX1}
\frac{N_X^2}{\alpha_X} = \frac{6}{\alpha_{\psi}} + \frac{2}{3\alpha_4 + 2\alpha_{2R}}.
\end{equation}
The boundary condition in Eq.\ref{eq:alphaX1}
applies at the symmetry breaking scale \(M_{GUT}\).

From Eq.\ref{Y} one finds the following relation between the
hypercharge gauge coupling constant \(g_Y\) and the \(SU(4)\) and
\(SU(2)_R\) gauge coupling constants \(g_4\) and \(g_{2R}\)
respectively:
\begin{equation} \label{eq:relation}
\frac{1}{\alpha_Y} = \frac{1}{\alpha_{2R}} + \frac{1}{\frac{3}{2}
\alpha_{4}}
\end{equation}
where \(\alpha_Y \equiv \frac{g^2_Y}{4\pi},~\alpha_{2R} \equiv
\frac{g^2_{2R}}{4\pi}\) and \(\alpha_4 \equiv \frac{g^2_4}{4\pi}\).
In terms of the ``GUT'' (in this case $E_6$)
normalized hypercharge generator $T_Y=\sqrt{\frac{5}{3}}Y$,
the coupling constant is
\(g_1 \equiv \sqrt{\frac{3}{5}} g_Y\):
\begin{equation} \label{eq:GUTrelation}
\frac{5}{\alpha_1} = \frac{3}{\alpha_{2R}} + \frac{2}{\alpha_{4}}
\end{equation}
where \(\alpha_1 \equiv \frac{g^2_1}{4\pi}\).
Eq.\ref{eq:GUTrelation} is a boundary condition for the gauge
couplings at the Pati-Salam symmetry breaking scale, in this case
$M_{GUT}$. Due to the left-right symmetry \(D_{LR}\), at the \(G_{4221}\) symmetry
breaking scale we have the additional boundary condition
$\alpha_{2L}=\alpha_{2R}$.

\begin{table}
\begin{center}
 \begin{tabular}{ | c | c |c | c| c| c| c | c| c| c | c| c | }
 \hline
 & \(Q\) & \(L\) &
 \(u^c\) & \(d^c\) &
 \(e^c\) & \(\nu^c\) & \(h_2\)
 & \(h_1\) & \(D\) &
 \(\overline{D}\) & \(S\)
 \\

\hline \(T_{B-L}\) &\(\frac{1}{6}\) & \(-\frac{1}{2}\) &
\(-\frac{1}{6}\) & \(-\frac{1}{6}\) & \(\frac{1}{2}\)& \(\frac{1}{2}\) & \(0\)
&
\(0\) & \(-\frac{1}{3}\) & \(\frac{1}{3}\) & \(0\) \\

\hline \(T^3_R\) &\(0\) & \(0\) &
\(-\frac{1}{2}\) & \(\frac{1}{2}\) & \(\frac{1}{2}\)& \(-\frac{1}{2}\) & \(\frac{1}{2}\)
&
\(-\frac{1}{2}\) & \(0\) & \(0\) & \(0\) \\

\hline \(T_{\psi}\) &\(\frac{1}{2}\) & \(\frac{1}{2}\) &
\(\frac{1}{2}\) & \(\frac{1}{2}\) & \(\frac{1}{2}\)& \(\frac{1}{2}\) & \(-1\)
&
\(-1\) & \(-1\) & \(-1\) & \(2\) \\

\hline \(Y\) &\(\frac{1}{6}\) & \(-\frac{1}{2}\) &
\(-\frac{2}{3}\) & \(\frac{1}{3}\) & \(1\)& \(0\) & \(\frac{1}{2}\)
&
\(-\frac{1}{2}\) & \(-\frac{1}{3}\) & \(\frac{1}{3}\) & \(0\) \\

\hline \(T_{\psi} + T^3_R\)&
\(\frac{1}{2}\)&\(\frac{1}{2}\)&\(0\)&\(1\)&\(1\)&\(0\)&\(-\frac{1}{2}\)&\(-\frac{3}{2}\)&\(-1\)&\(-1\)&\(2\)
\\ \hline
\end{tabular}
\end{center}
\caption{ \footnotesize
This table lists the \(T_{B-L} \equiv \frac{B-L}{2}\),
\(T^3_R\), \(T_{\psi}\), hypercharge \(Y \equiv T_{B-L} + T^3_R\), and
\(T_{\psi} + T^3_R\) charge assignments for the \(G_{3211}\)
representations of the \(\mathbf{27}\) multiplet of \(E_6\)
\cite{Slansky:1981yr}.  The charge for the \(U(1)_X\) group is
dependent on \(X \equiv (T_{\psi} + T^3_R) - c^{2}_{12} Y\) where
\(c^{2}_{12}\) is the square of the cosine of the mixing angle
\(\theta_{12}\) given by \(\tan \theta_{12} \equiv \mathbf{g}_{2R} /
\mathbf{g}_{B-L}\).}
\end{table}

In Table 1 we list the values that the generators \(Y\), \(T_{B-L}\), \(T^3_R\),
\(T_{\psi}\) and \(T_{\psi} + T^3_R\) (and therefore \(X\))
take for the \(G_{3211}\) representations of the \(\mathbf{27}\)
multiplet. Note that both \(T_{\psi} + T^3_R\) and \(Y\) are zero for
\(\nu^c\) and therefore neither \(B_Y\) or \(B_{X}\) couple to
right-handed neutrinos. This is a consequence of Goldstone's theorem
since the right-handed neutrino comes from the same \(G_{4221}\)
representation as the Higgs boson component that gets a VEV to break
the symmetry.

The $U(1)_X$ associated with the preserved generator in Eq.\ref{X}
is an anomaly-free gauge group which plays the same role in solving
the $\mu$ problem as the $U(1)_N$ of the E$_6$SSM, since it allows the
coupling $Sh_uh_d$ that generates an effective $\mu$ term, while
forbidding $S^3$ and the $\mu h_uh_d$. $U(1)_X$ is broken by the $S$
singlet VEV near the TeV scale, yielding a physical $Z'$ which may
be observed at the LHC. We emphasize that this $Z'$ is distinct from
those usually considered in the literature based on linear
combinations of the $E_6$ subgroups \(U(1)_{\psi}\) and
\(U(1)_{\chi}\) since, in the ME$_6$SSM, \(U(1)_{\chi}\) is necessarily
broken at $M_p$. In particular the $Z'$ of the ME$_6$SSM based on
$U(1)_X$ and that of the E$_6$SSM based on $U(1)_N$ will have different
physical properties, which we will explore later.

\section{Two-Loop RGEs Analysis}

In this section we perform a SUSY two-loop RG analysis of the gauge
coupling constants, corresponding to the pattern of symmetry breaking
discussed in the previous section. According to our assumptions there are
three complete $\mathbf{27}$ SUSY representations of the gauge group
$E_6$ in the spectrum which survive down to low energies, but,
unlike the original E$_6$SSM, there are no additional $H',\overline{H}'$
states and so the gauge couplings are not expected to converge at
$M_{GUT}$. We therefore envisage the pattern of symmetry breaking
shown in Eq.\ref{eq:steps}. Above the \(G_{4221}\) symmetry
breaking scale $M_{GUT}$ we assume, in addition to the three
$\mathbf{27}$ representations, some \(G_{4221}\) symmetry breaking
Higgs sector.

Although a Higgs sector consisting of \(H_R\) and \(\overline{H}_R\)
is perfectly adequate for breaking the \(G_{4221}\) symmetry, it does not
satisfy $D_{LR}$. We must therefore also consider an extended Higgs
sector including their left-right symmetric partners. A minimal
left-right symmetric Higgs sector capable of breaking the \(G_{4221}\)
symmetry consists of the \(SO(10) \otimes U(1)_{\psi}\)
Higgs states \((\mathbf{16}_H)_{\frac{1}{2}}\)
and \((\overline{\mathbf{16}}_{H})_{-\frac{1}{2}}\),
where \((\mathbf{16}_H)_{\frac{1}{2}}=(4,2,1)_{\frac{1}{2}} +
(\overline{4}, 1, \overline{2})_{\frac{1}{2}}\)
and $(\overline{\mathbf{16}}_{H})_{-\frac{1}{2}}=
(\overline{4},\overline{2},1)_{-\frac{1}{2}}
+(4,1,2)_{-\frac{1}{2}} $, where the components which get VEVs are
$\overline{H}_R=(\overline{4},1,\overline{2})_{\frac{1}{2}}$ and
\(H_R=(4,1,2)_{-\frac{1}{2}}\). If complete \(E_6\) multiplets are
demanded in the entire theory below $M_p$, then the Pati-Salam
breaking Higgs sector at $M_{GUT}$ may be assumed to be
\(\mathbf{27}_H\) and \(\overline{\mathbf{27}}_{H}\).
Therefore, in our
analysis we shall consider two possible Higgs sectors which
contribute to the SUSY beta functions in the region between $M_{GUT}$
and $M_p$, namely either
\((\mathbf{16}_H)_{\frac{1}{2}}+(\overline{\mathbf{16}}_{H})_{-\frac{1}{2}} \)
or \(\mathbf{27}_H+\overline{\mathbf{27}}_{H}\), where it is
understood that only the \(G_{4221}\) gauge group exists in this
region, and these Higgs representations must be decomposed under the
\(G_{4221}\) gauge group.

We therefore investigate the running of the gauge coupling constants at
two-loops for an \(E_6\) theory that contains three complete
\(\mathbf{27}\) multiplets at low energies and either a
\((\mathbf{16}_H)_{\frac{1}{2}} + (\overline{\mathbf{16}}_H)_{-\frac{1}{2}}\) or a \(\mathbf{27}_H +
\overline{\mathbf{27}}_H\) above the \(G_{4221}\) symmetry breaking
scale.  The \(E_6\) symmetry is assumed to be broken to a left-right
symmetric \(G_{4221}\) gauge symmetry which is then broken to the
standard model gauge group and a \(U(1)_X\) group as described in
section 2.  In the previous section we discussed the relation in
Eq.\ref{eq:GUTrelation} between the hypercharge and Pati-Salam gauge
coupling constants at the \(G_{4221}\) symmetry breaking scale. This
can be turned into a boundary condition involving purely Standard Model
gauge couplings constants at the \(G_{4221}\) breaking scale, since
\(SU(3)_c\) comes from \(SU(4)\) so \(\alpha_3 = \alpha_4\) at this
scale, and, as remarked, the \(D_{LR}\) symmetry requires that
\(\alpha_{2R} =\alpha_{2L}\) at the \(G_{4221}\) symmetry breaking
scale. Therefore Eq.\eqref{eq:GUTrelation} can be re-expressed as:
\begin{equation} \label{eq:GUTrelation2}
\frac{5}{\alpha_1} = \frac{3}{\alpha_{2L}} + \frac{2}{\alpha_{3}}
\end{equation}
Having specified the low energy matter content and thresholds,
Eq.\ref{eq:GUTrelation2} allows a unique determination of the
Pati-Salam breaking scale, by running up the gauge couplings
until the condition in Eq.\ref{eq:GUTrelation2} is satisfied.
In practice we determine the Pati-Salam symmetry breaking
scale to be close to the conventional
GUT energy scale, and this justifies our use of the notation $M_{GUT}$ to
denote the Pati-Salam breaking scale.  \begin{figure}
\includegraphics[angle=0, scale=1.012]{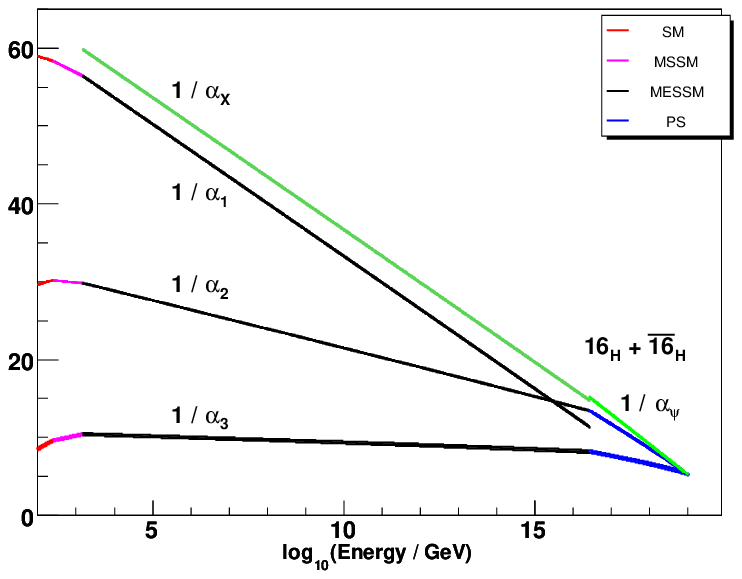} 
  \hspace{0.28 cm}
  \includegraphics[angle=0, scale=1.012]{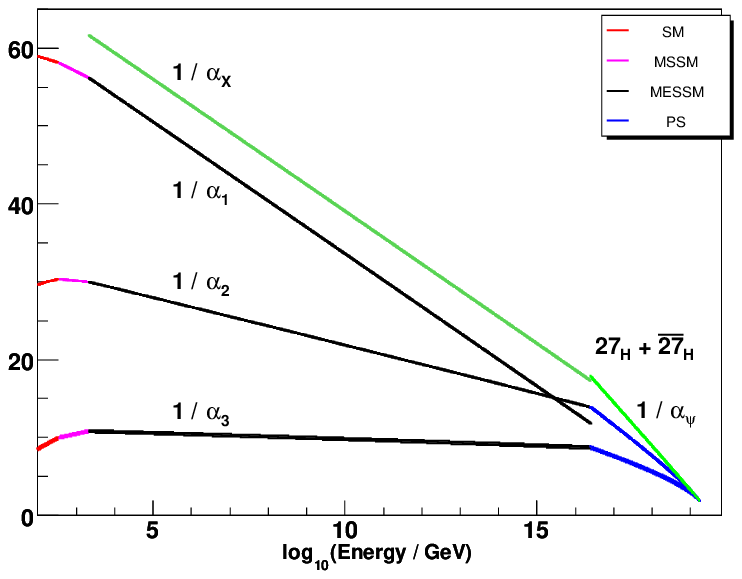} 
  \caption{\footnotesize This figure illustrates the two-loop RGEs
  running of the gauge coupling constants of two different \(E_6\)
  GUTs with an intermediate left-right symmetric \(G_{4221}\)
  symmetry.  Both models contain three \(\mathbf{27}\) supermultiplets
  of \(E_6\) at low energies which contain all the MSSM states as well
  as new (non-MSSM) states.  For the \(E_6\) model in the left-panel
  we assume effective MSSM and non-MSSM thresholds of \(250\) GeV and
  \(1.5\) TeV respectively.  For the \(E_6\) model in the right-panel
  we assume slightly larger effective MSSM and non-MSSM thresholds of
  \(350\) GeV and \(2.1\) TeV respectively.  Above the \(G_{4221}\)
  symmetry breaking scale we include the effects of \(SO(10) \otimes U(1)_{\psi}\)
  supermultiplets \((\mathbf{16}_H)_{\frac{1}{2}} + (\overline{\mathbf{16}}_H)_{-\frac{1}{2}}\) or
  \(E_6\) supermultiplets \(\mathbf{27}_H + \overline{\mathbf{27}}_H\)
  for the left-panel and right-panel respectively.  From the two-loop
  RGE analysis we find that the left-right symmetric \(G_{4221}\)
  symmetry is broken at \(10^{16.45(3)}\) GeV or \(10^{16.40(3)}\) GeV
  and that gauge coupling unification occurs at \(10^{18.95(8)}\) GeV
  or \(10^{19.10(10)}\) GeV for the left-panel and right-panel
  respectively.  For both panels we calculate that \(c^2_{12} \equiv
  \cos \theta_{12}\) is equal to \(0.71\) to two significant figures
  where \(\tan \theta_{12} \equiv g_{2R} / g_{B -
  L}\) and the gauge
  coupling constants are evaluated at the \(G_{4221}\) symmetry breaking scale.  The
  \(U(1)_X\) charge is given by \( X / N_X\) where \(X =
  (T^3_R + T_{\psi}) - c^2_{12} Y\) and \(N^2_X = 7 - 2 c^2_{12}
  + \frac{5}{3} c^4_{12}\).}
   \label{b}
\end{figure}
Above the scale $M_{GUT}$ we run up the two Pati-Salam gauge
couplings, namely $\alpha_4$ and $\alpha_{2L}=\alpha_{2R}$,
including, in addition to the three SUSY $\mathbf{27}$ matter
representations, also a Pati-Salam SUSY Higgs breaking sector
consisting of either \((\mathbf{16}_H)_{\frac{1}{2}}+(\overline{\mathbf{16}}_{H})_{-\frac{1}{2}}\) or
\(\mathbf{27}_H+\overline{\mathbf{27}}_{H}\).
The values of the gauge coupling constants meet at a high energy scale
close to the Planck scale $M_p=1.2\times 10^{19}$ GeV, which suggests
a Planck scale unification of all forces with gravity.
Assuming full gauge unification at the high energy scale,
the $U(1)_{\psi}$ gauge coupling is then run down,
along with the Pati-Salam gauge couplings, and the
$U(1)_{X}$ gauge coupling is determined at the \(G_{4221}\) symmetry
breaking scale $M_{GUT}$ from the boundary condition in Eq.\ref{eq:alphaX}.
The low energy gauge couplings, including $g_X$, are then run up
and the procedure is repeated until self-consistent unification
is achieved.

The results are shown in Figure 1.
For the \(E_6\) theory that contains the \((\mathbf{16}_H)_{\frac{1}{2}}+(\overline{\mathbf{16}}_{H})_{-\frac{1}{2}}\) particles, we take a low energy effective threshold
of \(250\) GeV for the MSSM states and therefore an effective
threshold of (\(6 \times 250\)) \(= 1.5\) TeV for the rest of states
of the three complete \(\mathbf{27}\) multiplets as assumed in
\cite{Howl:2007hq}, which follows the analysis of effective MSSM
thresholds from \cite{King:2007uj}.  For the \(E_6\) theory that
contains the \(\mathbf{27}_H + \overline{\mathbf{27}}_H\) particles,
the MSSM threshold must be increased to \(350\) GeV (and hence the
\(1.5\) TeV threshold is increased to \(2.1\) TeV) to ensure
unification for the gauge coupling constants of the \(G_{4221}\)
symmetry. We run the gauge couplings up from low
energies to high energies, using as input the SM gauge coupling constants measured
on the Z-pole at LEP, which are as follows \cite{Yao:2006px}:
\(\alpha_1 (M_Z) = 0.016947(6)\), \(\alpha_2 (M_Z) = 0.033813(27)\)
and \(\alpha_3(M_Z) = 0.1187(20)\). The general two-loop beta
functions used to run the gauge couplings can be found in
\cite{two-loop}. Using a two-loop renormalization group analysis, we
calculate that the \(G_{4221}\) symmetry is broken at
\(10^{16.45(3)}\) GeV or \(10^{16.40(3)}\) GeV and that gauge coupling
unification occurs at \(10^{18.95(8)}\) GeV or \(10^{19.10(10)}\) GeV
for the \(E_6\) theories that contain \((\mathbf{16}_H)_{\frac{1}{2}}+(\overline{\mathbf{16}}_{H})_{-\frac{1}{2}}\) or \(\mathbf{27}_H +
\overline{\mathbf{27}}_H\) respectively.
In \cite{Howl:2007hq} the same two-loop RGE analysis was carried out
for the \(E_6\) model but without the inclusion of the \(U(1)_{\psi}\)
and \(U(1)_X\) groups.  In terms of a logarithmic scale, the
Pati-Salam symmetry breaking scale and unification scale have not been
significantly changed from \cite{Howl:2007hq} by the inclusion of
\(U(1)_X\) and \(U(1)_{\psi}\); and Planck scale unification and GUT
scale Pati-Salam symmetry breaking is still a possibility.

\begin{table}
\small
\begin{center}
 \begin{tabular}{ | c | c |c | c| c| c| c | c| c| c | c| c | }
 \hline
 & \(Q\) & \(L\) &
 \(u^c\) & \(d^c\) &
 \(e^c\) & \(\nu^c\) & \(h_2\)
 & \(h_1\) & \(D\) &
 \(\overline{D}\) & \(S\)
  \\  \hline \(Y\) &\(\frac{1}{6}\) & -\(\frac{1}{2}\) & -\(\frac{2}{3}\) &
\(\frac{1}{3}\) & \(1\)& \(0\) & \(\frac{1}{2}\) &
-\(\frac{1}{2}\) & -\(\frac{1}{3}\) & \(\frac{1}{3}\) & \(0\) \\
\hline \(X\)& \(
\frac{8}{21}\)&\(\frac{6}{7}\)&\(
\frac{10}{21}\)&\(\frac{16}{21}\)&\(\frac{2}{7}\)&\(0\)&-\(\frac{6}{7}\)&-\(\frac{8}{7}\)&-\(\frac{16}{21}\)&-\(
 \frac{26}{21}\)&\(2\) \\ \hline \(N\)& 1&2&1&2&1&0&-2&-3&-2&-3&5 \\
 \hline
 \(T_Y\) &\(0.129\) & -\(0.387\) & -\(0.516\) &
\(0.258\) & \(0.775\)& \(0\) & \(0.387\) &
-\(0.387\) & -\(0.258\) & \(0.258\) & \(0\) \\
\hline \(T_X\)& \(
0.150\)&\(0.338\)&\(
0.188\)&\(0.301\)&\(0.113\)&\(0\)&-\(0.338\)&-\(0.451\)&-\(0.301\)&-\(
 0.489\)&\(0.789\) \\ \hline \(T_N\)& 0.158  &   0.316  &  0.158  &  0.316  &  0.158  &  0  &  -0.316&  -0.474&  -0.316&  -0.474&  0.791 \\
 \hline
\end{tabular}
\end{center}
\caption{\footnotesize This table lists the values that the charges
\(Y\), \(X\) and \(N\) take for the all the \(G_{3211}\)
representations of the \(E_6\) \(\mathbf{27}\) multiplet.  \(Y\) is
hypercharge, \(X\) is the charge of \(U(1)_X\) for the model presented
in sections 3 and 4 and \(N\) is the charge associated with the
\(U(1)_N\) group in the E$_6$SSM.  These charges are normalized by the
\(E_6\) normalization constants \(N^2_Y \equiv \frac{5}{3}\), \(N^2_X
= 6 \frac{62}{147}\) and \(N^2_N \equiv 40\) so that the \(E_6\)
normalized charges of the \(U(1)_Y\), \(U(1)_X\) and \(U(1)_N\) groups
are given by \(T_Y \equiv Y / N_Y\), \(T_X \equiv X / N_X\) and \(T_N
\equiv N / N_N\) respectively, which are also given in the table.
\(N_X\) and \(X\) have been calculated for the case when the mixing
angle \(\tan \theta_{12} = g_{2R} / g_{B-L}\) is
given by \(\cos^2 \theta_{12} \equiv c^2_{12} = 5 / 7\) which, to two
significant figures, agrees with the two-loop RGEs analysis presented
for the two \(E_6\) theories in section 3.}
\end{table}

The value of the gauge coupling constant at the unification scales
\(10^{18.95(8)}\) GeV or \(10^{19.10(10)}\) GeV is \(\alpha_P =
0.183(10)\) or \(\alpha_P = 0.432(121)\) for the \((\mathbf{16}_H)_{\frac{1}{2}}+(\overline{\mathbf{16}}_{H})_{-\frac{1}{2}}\) or \(\mathbf{27}_H +
\overline{\mathbf{27}}_H\) particle spectra, respectively.
The values of the unified gauge coupling at the Planck scale are
much larger than the conventional values of $\alpha_{GUT}$ and
indeed are larger even than $\alpha_3(M_Z)$, however they are still
in the perturbative regime. Of course there are expected to be large
threshold corrections coming from Planck scale physics which are not
included in our analysis. Indeed, we would expect that QFT breaks
down as we approach the Planck scale, so that our RG analysis ceases
to be valid as we approach the Planck scale, as remarked in the
Introduction. The precise energy scale \(E_{new}\) at which quantum field
theories of gravity are expected to break down and new physics takes
over is discussed in \cite{Han:2004wt} based on estimates of the
scale of violation of (tree-level) unitarity.
An upper bound for this new physics energy scale
is given by \(E^2_{new} = 20 [G (\frac{2}{3} N_s + N_f + 4N_V)]^{-1}
\) where \(N_s\), \(N_f\) and \(N_V\) are the number of scalars,
fermions and vectors respectively that gravity couples to. Assuming three
low-energy \(\mathbf{27}\) multiplets, \(E_{new}\) would be equal to
\(10^{18.6}\) GeV which sets an upper bound for the scale at which
our quantum field theory analysis (and with any corrections from
effective quantum gravity theories included) can no longer be
trusted. We have shown that the gauge coupling constants are
predicted to be very close to one another at this scale and that, if
extrapolated, unify just below \(M_p\). We have naively
extrapolated the RGEs up to \(M_p\), even though new physics
associated with quantum gravity must enter an order of magnitude
below this. The fact that the two PS couplings are
very close to each other at \(E_{new}\), and are on a
convergent trajectory must be regarded, at best, as a suggestive hint of a
unification of the gauge fields with gravity in this approach.

From the previous section, the charge of the \(U(1)_X\) group \(T_X\)
 depends on the values that the \(g_4\) and \(g_{2R}\) coupling
 constants take at the \(G_{4221}\) symmetry breaking scale, which is
 written into the cosine \(c_{12}\) of the mixing angle \(\tan
 \theta_{12} \equiv g_{2R} / g_{B - L}\).  The value
 that \(g_4\) and \(g_{2R}\) take at this scale is automatically set
 by the unification of the \(G_{4221}\) gauge coupling constants.  We
 calculate that, with either \((\mathbf{16}_H)_{\frac{1}{2}}+(\overline{\mathbf{16}}_{H})_{-\frac{1}{2}}\) or \(\mathbf{27}_H +
 \overline{\mathbf{27}}_{H}\) included above the \(G_{4221}\)
 symmetry breaking scale, \(c^2_{12}\) is equal \(0.71\) to two
 significant figures.  However, for convenience we take the physical
 value of \(c^2_{12}\) to be equal to \(\frac{5}{7}\) (\(\sim 0.71\))
 so that \(T_X\) can be written in terms of fractions.  Using this
 value of \(c^2_{12}\) in equation Eq.\ref{eq:Z}, we can calculate
 \(T_X\) for all the standard model particles of the three low-energy
 \(\mathbf{27}\) multiplets. The values that \(T_X\), \(T_Y\) and
 \(T_N\) take for the particles of the \(\mathbf{27}\) multiplets are
 given in Table 2, where \(T_N\) is the generator associated with the
 \(U(1)_N\) group in the E$_6$SSM.

\section{Phenomenology of the new $Z'$ in the ME$_6$SSM}

In this section we investigate certain phenological implications of the $Z'$ gauge boson in the ME$_6$SSM.  We compare the results to those calculated for the $Z'$ in the E$_6$SSM to see if a possible distinction could be made between the two models in future experiments.  We start by reviewing the covariant derivatives for the E$_6$SSM and ME$_6$SSM symmetries below the GUT scale and compare the different $U(1)'$ groups from the two models.  The mixing between the $Z'$ of the ME$_6$SSM and the Standard Model $Z$ gauge boson is then calculated and shown to be negligible as in the E$_6$SSM. We then calculate the axial and vector couplings of the $Z'$ to the low energy particle spectrum and show that the charged lepton vector couplings do differ in the E$_6$SSM and ME$_6$SSM, which could potentially lead to a distinction between the two models in future experiments.

\subsection{The $Z'$ of the E$_6$SSM}

In the E$_6$SSM the \(E_6\) symmetry is not broken through a Pati-Salam
intermediate symmetry but instead breaks to \(SU(3)_c \otimes SU(2)_L
\otimes U(1)_Y \otimes U(1)_N\) via a \(E_6 \rightarrow SO(10) \otimes
U(1)_{\psi} \rightarrow SU(5) \otimes U(1)_{\chi} \otimes
U(1)_{\psi}\) symmetry breaking chain.  The covariant derivative for
the \(SU(3)_c \otimes SU(2)_L \otimes U(1)_Y \otimes U(1)_N\) symmetry
can be written as:

\begin{equation} \label{eq:E$_6$SSM}
D_{\mu} = \partial_{\mu} + ig_{3} T^n_{3c} A^n_{3c \mu} +
ig_{2L} T^s_{L} A^s_{L \mu} + i g_1 T_Y B_{Y \mu} + i g_N T_N  B_{N \mu}
\end{equation}

where \(n = 1 \ldots 8\) and \(s = 1 \ldots 3\);  \(A^{n}_{3c\mu}\), \(A^s_{L \mu}\), \(B_{Y \mu}\) and \(B_{N \mu}\) are the \(SU(3)_c\), \(SU(2)_L\),
\(U(1)_Y\) and \(U(1)_{N}\) quantum fields respectively;
\(g_{3}\), \(g_{2L}\), \(g_{1}\) and \(g_{N}\) denote the universal gauge coupling
constants of the respective fields and \(T^n_{3c}\), \(T^s_L\), \(T_Y\) and
\(T_{N}\) represent their generators.
At low energies the $U(1)_N$ gauge group will be broken,
giving rise to a massive $Z'$ gauge boson associated with the 
E$_6$SSM.

The \(g_N\) gauge coupling constant is equal to \(g_1\) to an excellent
approximation \cite{E$_6$SSM}, independent of the energy scale of
interest.  This is to be compared to the universal gauge coupling constant
\(g_X\) of the group \(U(1)_X\) in the models presented in
this section, which is always less than \(g_1\).

Similar to \(T_Y\) and \(T_X\), we can split \(T_N\) into an \(E_6\)
normalization constant \(N_N\) and a non-normalized charge \(N\) so
that \(T_N \equiv N / N_N\) where the conventional choice is \(N^2_N \equiv
40\) and \(N \equiv \frac{1}{4} \chi + \frac{5}{2} T_{\psi}\) where
\(\chi \equiv 2 \sqrt{10} T_{\chi}\) \cite{E$_6$SSM}.

\subsection{The $Z'$ of the ME$_6$SSM}

The covariant derivative of the \(G_{4221}\)
symmetry is discussed in the Appendix and is given by Eq.\ref{eq:covariant derivative} as:

\[
D_{\mu} = \partial_{\mu}  + ig_4 T_4^m A^m_{4 \mu} + ig_{2L} T^s_L
A^s_{L \mu} + ig_{2R} T^r_R
A^r_{R \mu} +  \frac{1}{\sqrt{6}} i g_{\psi} T_{\psi} A_{\psi \mu}
\]

where \(m = 1 \ldots 15\) and \(r,s = 1 \ldots 3\);  \(A^{m}_{4
\mu}\), \(A^r_{R \mu}\) and \(A_{\psi \mu}\) are the \(SU(4)\),
\(SU(2)_R\) and \(U(1)_{\psi}\) quantum fields respectively;
\(g_4\), \(g_{2R}\) and \(g_{\psi}\) denote the universal gauge coupling
constants of the respective fields; and \(T^m_4\), \(T^r_R\) and
\(T_{\psi}\) represent their generators.

The covariant derivative of the \(G_{3211}\) symmetry is derived in the Appendix and is given by Eq.\ref{eq:DmuG3211} as:

\begin{equation} \label{eq:DmuG3211a}
D_{\mu} = \partial_{\mu} + ig_{3} T^n_{3c} A^n_{3c \mu} +
ig_{2L} T^s_{L} A^s_{L \mu} + i g_1 T_Y B_{Y \mu} + i g_X T_X B_{X \mu}
\end{equation}

where \(n = 1 \ldots 8\) and \(s = 1 \ldots 3\);  and \(B_{X \mu}\)
and \(T_X\) are the gauge field of the \(U(1)_X\) group and its
(\(E_6\) normalized) charge respectively.
At low energies the $U(1)_X$ gauge group will be broken,
giving rise to a massive $Z'$ gauge boson associated with the 
ME$_6$SSM.

As is clear from Table 2,
for \(c^2_{12} = \frac{5}{7}\), the \(T_X\) and \(T_N\) charges are
different for all of the \(G_{3211}\) representations of the
\(\mathbf{27}\) multiplets.
However, in the limit \(c^2_{12} = \frac{3}{5}\),
corresponding to \(g_{2R}= g_4\) at the \(G_{4221}\) symmetry breaking scale,
then \(T_X\) and \(T_N\) are identical.\footnote{Although \(T_X\) and
\(T_N\) are identical for \(c^2_{12} = 3 / 5\), \(X\) and \(N\) and
hence \(N_X\) and \(N_N\) are not.  However, we could have defined
\(X\) and \(N_X\) differently so that they agree with \(N\) and
\(N_N\) when \(c^2_{12} = 3 / 5\).}  This can be seen if one sets
\(g_{2R} = g_4 = \sqrt{\frac{2}{3}} g_{B-L}\)
in Eq.\ref{X} and Eq.\ref{eq:NX}, in which case \(T_X\) is given by:

\[
T_X = \frac{1}{4} \bigg[  \frac{4}{2 \sqrt{10}} \Big(T^3_R - \frac{3}{2} T_{B - L} \Big) + \sqrt{15} \Big(T_{\psi} / \sqrt{6} \Big) \bigg]\]

\[~~~~\equiv \frac{1}{4} \bigg[   T_{\chi} + \sqrt{15} \Big(T_{\psi} / \sqrt{6} \Big) \bigg]\]

\[~~~~\equiv T_{\chi} \cos \theta + (T_{\psi} / \sqrt{6}) \sin \theta\]

\[~~~~\equiv T_N ~~\mathrm{(see~ [14])}\] 

where \(\theta = \arctan \sqrt{15}\) and \(T_{\chi}\) is the \(E_6\)
normalized charge for the \(U(1)_{\chi}\) group, which is defined by
\(SO(10) \rightarrow SU(5) \otimes U(1)_{\chi}\)
\cite{Keith:1997zb}
\footnote{When \(g_{2L}=  g_{2R} = g_4\) the Pati-Salam generators
can be thought of as $SO(10)$ generators, on the same footing as
the $SU(5)$ and $U(1)_{\chi}$ generators when their gauge couplings
are equal, as in the E$_6$SSM.
In this limit the above argument shows that
there is no distinction between $U(1)_N$ and
$U(1)_X$.}.
In the E$_6$SSM the \(U(1)_N\) group is defined as the linear
combination of the two groups \(U(1)_{\chi}\) and \(U(1)_{\psi}\) for
which the right-handed neutrino is a singlet of the symmetry
\cite{E$_6$SSM}.  This linear combination is \(U(1)_{\chi} \cos \theta +
U(1)_{\psi} \sin \theta\), where \(\theta = \arctan \sqrt{15}\)
\cite{E$_6$SSM}, which is the same linear combination of \(U(1)_{\chi}\)
and \(U(1)_{\psi}\) that \(U(1)_X\) becomes if \(g_R =
g_4\) as shown above.
Thus if \(g_R =
g_4\) at the \(G_{4221}\) symmetry breaking scale, then the
covariant derivative for the E$_6$SSM, Eq.\ref{eq:E$_6$SSM}, and the covariant
derivative for \(G_{3211}\), Eq.\ref{eq:DmuG3211a}, become equivalent
because of the reasons stated above.
However, in the \(E_6\) theories that
we have proposed, \(c^2_{12} \sim \frac{5}{7}\) not \(\frac{3}{5}\) so
that, in general, one expects
\(g_R \neq g_4\) at the \(G_{4221}\) symmetry breaking scale in realistic models.

It is clearly of interest to try to distinguish the
$Z'$ of the E$_6$SSM from that of the ME$_6$SSM, since the former one is
associated with GUT scale unification, while the latter is
associated with Planck scale unification. In the remainder
of this section we discuss the phenomenology of the
new  $Z'$ of the ME$_6$SSM, comparing it to that of the E$_6$SSM.
In principle different $Z'$ gauge bosons can be distinguished
at the LHC by measuring the leptonic forward-backward charge
asymmetries as discussed in \cite{Dittmar:2003ir}, providing the
mass of the $Z'$ is not much larger than about 2 TeV.

\subsection{Mixing between \(Z\) and the \(Z'\) of the ME$_6$SSM}
In this section we investigate the mixing between the \(Z\) gauge boson and the \(Z'\) gauge
boson of \(U(1)_X\) which is generated once the MSSM Higgs doublets get vacuum expectation values and break the electroweak symmetry.  When the MSSM singlet particle \(S\) from the low-energy \(\mathbf{27}\) multiplets of the ME$_6$SSM gets a VEV, the \(U(1)_X\) group will be broken and a heavy \(Z'\) gauge boson will be produced.  Then, when \(h_u\) and \(h_d\) get VEVs, the \(SU(2)_L \otimes U(1)_Y\) symmetry will be broken to \(U(1)_{em}\) and a heavy neutral \(Z\) gauge boson, which is the following mixture of the \(SU(2)_L\) and \(U(1)_Y\) fields: \(Z_{\mu} = W^3_{\mu} \cos \theta_W - A_Y \sin
\theta_W\) where \(\theta_W\) is the Electroweak (EW) symmetry mixing angle.   Since \(h_u\) and \(h_d\) transform under \(U(1)_X\), they couple to \(Z'\) and so mix the \(Z'\) and \(Z\) gauge bosons when they get VEVs.  After \(S\), \(h_u\) and \(h_d\) get VEVs the mass squared mixing matrix for the \(Z\) and \(Z'\) gauge bosons is given by \cite{Babu,Kang:2004bz}:

\[M^2_{Z Z'} =  \left ( \begin{array} {cc} M^2_Z & \delta M^2  \\
\delta M^2  & M^2_{Z'}
\end{array} \right )\]

where:

\[M^2_Z = (g^2_{2L} + g^2_Y ) (Y^h)^2 v_h^2\]
\[M^2_{Z'} = g^2_X v_h^2 [ (T^{h_1}_X)^2 \cos^2 \beta + (T^{h_2}_X)^2 \sin^2 \beta ] + g^2_X (T^S_X)^2 s^2\]
\[\delta M^2 = \sqrt{g^2_{2L} + g^2_Y}~  g_X~ Y^h (T^{h_1}_X \cos^2 \beta - T^{h_2}_X \sin^2 \beta) v_h^2\]

where \(Y^h\) is the magnitude of the \(h_u\) and \(h_d\) Higgs bosons' hypercharge; \(T^{h_1}_X\), \(T^{h_2}_X\) and \(T^S_X\) are the values that the \(E_6\) normalized \(U(1)_X\) charge, \(T_X\), takes for the \(h_1\), \(h_2\) and \(S\) states respectively;  \(g_{2L}\) and
\(g_Y\) are the \(SU(2)_L\) and (non-GUT normalized) hypercharge gauge coupling
constants evaluated at the EW symmetry breaking scale;\footnote{The non-GUT normalized hypercharge coupling constant \(g_Y\) is identified as \(g_Y \equiv \sqrt{\frac{3}{5}} g_1\).}
\(g_X\) is the \(U(1)_X\) gauge coupling constant evaluated at
the \(U(1)_X\) symmetry breaking scale; \(s\) is the VEV of the MSSM singlet \(S\); \(v_h = \sqrt{v^2_u + v^2_d}\) and \(\tan \beta = \frac{v_d}{v_u}\) where \(v_u\) and \(v_d\) are the vacuum expectation values for the \(h_u\) and \(h_d\) MSSM Higgs bosons respectively.

The mass eigenstates generated by this mass mixing matrix are:

\[Z_1 = Z \cos \theta_{Z Z'} + Z' \sin \theta_{Z Z'}\]
\[Z_2 = - Z \sin \theta_{Z Z'} + Z' \cos \theta_{Z Z'}\]

with masses \(M^2_{Z_1, Z_2} = \frac{1}{2} [ M^2_Z + M^2_{Z'} \mp
\sqrt{(M^2_Z - M^2_{Z'}) + 4 \delta M^4} ] \) respectively.  The mixing angle \(\theta_{Z Z'}\) is given by:

\[\tan ( 2 \theta_{Z Z'}) = \frac{2 \delta M^2}{M^2_{Z'} - M^2_Z}.\]

In terms of this mixing angle the covariant derivative for the mass eigenstate gauge bosons \(Z_1\) and \(Z_2\) is:

\[D_{\mu} = \partial_{\mu} + i \bigg( \frac{ \cos \theta_{Z Z'} }{\sqrt{ g^2_Y + g^2_{2L}}} ( g^2_{2L} T^3_L - g^2_Y Y) - g_X T_X \sin \theta_{Z Z'} \bigg) Z_{1 \mu}\]

\[~~~~~~~~~~~+ i \bigg( g_X T_X \cos \theta_{Z Z'} + \frac{ \sin \theta_{Z Z'}}{\sqrt{g^2_Y + g^2_{2L}}} ( g^2_{2L} - g^2_Y Y ) \bigg) Z_{2 \mu}
\]

where \(g_Y\) and \(g_{2L}\) are evaluated at the EW symmetry breaking scale and \(g_X\) is evaluated at the scale at which \(S\) gets a VEV to break the \(U(1)_X\) symmetry.  Phenomenology constrains the mixing angle \(\theta_{Z Z'}\) to be typically less than \(2 - 3 \times 10^{-3}\) \cite{Abreu:2000ap} and the mass of the extra neutral gauge boson to be heavier than \(500 - 600\) GeV \cite{Abe:1997fd}.  We calculate that, if the \(S\) particle gets a VEV at \(1.5\) TeV in the ME$_6$SSM, then \(\theta_{Z Z'} = 3 \times 10^{-3}\) and \(M_{Z'} = 544\) GeV so that phenomenologically  acceptable values are therefore produced for \(s > 1.5\) TeV.  This vacuum expectation value is consistent with the RGEs analysis in section 3 and the scale of electroweak symmetry breaking.

Since the mixing angle \(\theta_{Z Z'}\) is very small in the ME$_6$SSM, we approximate the two mass eigenstate gauge bosons to be just \(Z\) and \(Z'\), which are the neutral gauge bosons of the broken \(SU(2)_L \otimes U(1)_Y\) and \(U(1)_X\) symmetries respectively.  The above covariant derivative is then simplified to:

\[D_{\mu} = \partial_{\mu} + i \frac{1}{\sqrt{g^2_Y + g^2_{2L}}} Z_{\mu} \big( g^2_{2L} T^3_{L} - g^2_Y Y \big) + i g_X Z_{\mu}'  T_X
.\]

\subsection{Axial and Vector Couplings for \(Z'\) in the ME$_6$SSM}

If we ignore the mixing between the \(Z\) and \(Z'\) gauge bosons, then the most general Lagrangian for the \(U(1)_X\) group is \cite{Babu,Babu:1996vt}:

 \[
 \mathcal{L}_X = \frac{1}{2} M_{Z'} Z'^{\mu} Z'_{\mu} - \frac{g_X}{2} \sum_i \overline{\psi}_i \gamma^{\mu} (f^i_V - f^i_A \gamma^5) \psi_i Z'_{\mu} - \frac{1}{4} F'^{\mu \nu} F'_{\mu \nu}  - \frac{\sin \chi}{2} F'^{\mu \nu} F_{\mu \nu}
 \]

\begin{table}
\begin{center}
 \begin{tabular}{ | c | c |c | c| c| c| c | c|}
 \hline
 & \(u\) & \(d\) &
 \(e\) & \(\nu\) &
 \(D\) & \(h\) & \(S\)
  \\ \hline \(f_V / N_X \)
&\(\frac{1}{2} - \frac{5}{6} c^2_{12}\) & \(-\frac{1}{2} +
\frac{1}{6} c^2_{12}\) & \(-\frac{1}{2} + \frac{3}{2}
c^2_{12}\) & \(\frac{1}{2} + \frac{1}{2}
c^2_{12}\) & \(\frac{2}{3} c^2_{12}\)& \(-1 +
c^2_{12}\) & \(2\)
\\ \hline
\( f_A/ N_X\)& \(\frac{1}{2} + \frac{1}{2}
c^2_{12}\)&\(\frac{3}{2} - \frac{1}{2}
c^2_{12}\)&\(\frac{3}{2} - \frac{1}{2}
c^2_{12}\)&\(\frac{1}{2} + \frac{1}{2}
c^2_{12}\)&\(-2\)&\(-2\)&\(2\)
 \\
\hline \( f_V \) 
&
\(-0.0376\)&\(-0.1503\)&\(0.2255\)&\(0.3382\)&\(0.1879\)&\(-0.1127\)&\(0.7892\)
 \\
\hline \( f_A  \)&
\(0.3382\)&\(0.4510\)&\(0.4510\)&\(0.3382\)&\(-0.7892\)&\(-0.7892\)&\(0.7892\)
\\ \hline
\( f^0_V \)&
\(0.0278\)&\(-0.1637\)&\(0.1081\)&\(0.2996\)&\(0.1359\)&\(-0.1915\)&\(0.7906\)
\\ \hline
\( f^0_A  \)&
\(0.2996\)&\(0.4910\)&\(0.4910\)&\(0.2996\)&\(-0.7906\)&\(-0.7906\)&\(0.7906\)
\\ \hline
\end{tabular}
\end{center}
\caption{In this table we list the axial \(f_A\) and vector \(f_V\)
\(U(1)_X\) charge assignments for the \(G_{3211}\) representations of
the complete \(\mathbf{27}\) \(E_6\) multiplet in the ME$_6$SSM.
The assignments for a
general ME$_6$SSM model and for the model presented in
sections 3 and 5, which has \(c^2_{12} = 5/7\), are both given.  The
\(E_6\) normalization factor \(N_X\) is given by \(N^2_X = 7 - 2
c^2_{12} + \frac{5}{3} c^4_{12}\) for a general model and is equal to
\(6 \frac{62}{147}\) when \(c^2_{12} = 5/7\).  We have also included
the axial and vector \(U(1)_N\) charge assignments $f^0_V$
and $f^0_A$ in the E$_6$SSM
so that a comparison can be made to the
corresponding ME$_6$SSM quantities $f_V$
and $f_A$.}
\end{table}

where \(F'^{\mu \nu}\) and \(F^{\mu \nu}\) are the field strength
tensors for \(U(1)_X\) and \(U(1)_Y\) respectively; \(\psi_i\) are the
chiral fermions and \(f^i_V\) and \(f^i_A\) are their vector and axial
charges which are given by \(f^i_V \equiv \frac{1}{N_X} ( X^i_L +
X^i_R ) \) and \(f^i_A \equiv \frac{1}{N_X} ( X^i_L - X^i_R ) \) where
\(X_L\) and \(X_R\) are the \(X\) charges for the left-handed and
right-handed particles respectively.

The \(\frac{\sin \chi}{2} F'^{\mu \nu} F_{\mu \nu}\) term in the above
Lagrangian represents the kinetic term mixing for the two Abelian
symmetries \(U(1)_Y\) and \(U(1)_X\).  In general, the kinetic term
mixing of two Abelian gauge groups is non-zero because the field
strength tensor is gauge-invariant for an Abelian theory.  However, if
both Abelian groups come from a simple gauge group, such as \(E_6\),
then \(\sin \chi\) is equal to zero at the tree-level, although
non-zero elements could arise at higher orders if the trace of the
\(U(1)\) charges is not equal to zero for the states lighter than the
energy scale of interest \cite{Babu:1996vt}.  The trace of the
\(U(1)_Y\) and \(U(1)_X\) charges is given by:

\[\mathrm{Tr}~ ( T_Y ~T_X ) = \sum_{\mathrm{i = chiral~fields}} ( T^i_Y ~T^i_X ).\]

This trace is only non-zero if incomplete GUT multiplets are present
in the low energy particle spectrum.  There are no low-energy
incomplete \(E_6\) multiplets in the model we presented in section 4
and so \(\sin \chi = 0\) at the tree-level and at higher orders in
this particular case.  There is therefore no kinetic term mixing
between the \(U(1)_Y\) and \(U(1)_X\) groups in the ME$_6$SSM.  In the
E$_6$SSM two additional EW doublets from incomplete \(E_6\) multiplets
are kept light so that, in this case, \(\sin \chi\) is non-zero
\cite{E$_6$SSM}.

The second term in the \(U(1)_X\) Lagrangian \(\mathcal{L}_X\)
represents the interaction between the \(Z'\) gauge boson and the
fermions.  In table 3 we list the vector and axial \(U(1)_X\) charges
for the \(G_{4221}\) representations of the complete \(\mathbf{27}\)
low-energy \(E_6\) multiplets in a general \(E_6\) theory and the
ME$_6$SSM, which has \(c^2_{12} = 5 / 7\).  We also list the
vector and axial \(U(1)_N\) charges of the E$_6$SSM for the low-energy
\(\mathbf{27}\) multiplets for a comparison.  The differences between
the values of the vector and axial couplings of the two \(Z'\) gauge
bosons of the \(U(1)_X\) and \(U(1)_N\) groups are due to the
difference in value between the \(E_6\) normalized \(T_X\) and \(T_N\)
charges and the fact that the kinetic term mixing between the
\(U(1)_Y\) and the \(U(1)'\) groups is non-zero in the E$_6$SSM but
zero in the ME$_6$SSM.  The largest difference between the vector and
axial couplings of \(U(1)_X\) and \(U(1)_N\) exists for the charged
leptons where the vector coupling for \(U(1)_X\) is a factor of two
larger than for \(U(1)_N\).

\section{A Realistic Model}
In this section we construct a realistic ME$_6$SSM,
focussing on the model building issues.
The ME$_6$SSM has a more `minimal'
particle content than the E$_6$SSM since it only contains three
complete \(\mathbf{27}\) multiplets at low energies whereas the
E$_6$SSM contains two additional EW doublets which can be considered
as states of incomplete \(\mathbf{27}\) and \(\overline{\mathbf{27}}\)
\(E_6\) multiplets.
From the RGEs analysis, unification of the
\(G_{4221}\) gauge coupling constants
occurs near the Planck scale where an \(E_6\) symmetry
should in principle exist. However, given the expected strength of
quantum gravity at this scale, it is likely that any such \(E_6\)
symmetry is for all practical purposes broken by gravitational effects.
Therefore, the model that we
propose in this section is chosen to not respect an \(E_6\) symmetry
but instead obey the \(G_{4221}\) symmetry that exists between the
conventional GUT and Planck scales where quantum gravity effects will
not be so significant.

\begin{table}
\begin{center}
\begin{tabular}{|c|c|c|c|}
\hline
field & $SU(4)\otimes SU(2)_{L}\otimes SU(2)_{R}\otimes U(1)_{\psi}$ & $U(1)_R$  & $Z^H_{2}$\\
\hline
${F_i}$, $F^{c}_{i}$ &  $(4,2,1)_{\frac{1}{2}}$,
$(\overline{4},1,2)_{\frac{1}{2}}$ & $1$ & $-$\\
\hline
$h_3$, $h_{\alpha}$  & $(1,2,2)_{-1}$  & $0$ & $+,-$\\
\hline
$S_3$, $S_{\alpha}$  & $(1,1,1)_{2}$  & $2$ & $+,-$\\
\hline
$\mathcal{D}_i$  & $(6,1,1)_{-1}$  & $0$ & $-$\\
\hline
$M$, $\Sigma$   & $(1,1,1)_{0}$  & $0$ & $+,-$\\
\hline
$H_L$, $\overline{H}_R$  & $(4,2,1)_{\frac{1}{2}}$,
$(\overline{4},1,2)_{\frac{1}{2}}$ &
$2$ & $+$\\
\hline
$\overline{H}_L$, $H_R$   & $(\overline{4},2,1)_{-\frac{1}{2}}$,
$({4},1,2)_{-\frac{1}{2}}$ & $0$ & $+$\\
\hline
\end{tabular}
\caption{\footnotesize
This table lists all the charge assignments
for the \(G_{4221}\) representations of the M$E_6SSM$,
where \(i = 1 \ldots 3\) is a family index and \(\alpha = 1,2\).
The \(U(1)_R\) is an R-symmetry and \(Z^H_2\) distinguishes
the third family Higgs which get VEVs.
These symmetries obey the
\(G_{4221}\) symmetry but not the \(E_6\) symmetry since the latter is
assumed to be broken by quantum gravity effects. The superpotential
terms that are allowed by these symmetries are given in Table 5.
The \(h_3\) supermultiplet contains the MSSM Higgs
bosons and \(S_3\) is the MSSM singlet that generates an effective
\(\mu\)-term \(\lambda_S S_3 h_3 h_3\).
The \(H_L\) and \(\overline{H}_L\)
Higgs bosons are required to satisfy the left-right discrete operator
\(D_{LR}\) that is defined by \(E_6 \rightarrow G_{4221} \otimes
D_{LR}\). \(\Sigma\) and \(M\) are \(E_6\) singlets
that are assumed to get VEVs at \(10^{7-11}\) GeV and \(10^{16.4}\) GeV
respectively.}
\end{center}
\end{table}

Under \(E_6 \rightarrow SO(10) \otimes U(1)_{\psi} \rightarrow
G_{4221}\), the fundamental \(E_6\) representation breaks into the
following: \(\mathbf{27} \rightarrow 16_{\frac{1}{2}} + 10_{-1} + 1_2
\rightarrow F + F^c + h + \mathcal{D} + S\) where \(F \equiv
(4,2,1)_{\frac{1}{2}}\) contains one family of the left-handed quarks
and leptons, \(F^c \equiv
(\overline{4},1,\overline{2})_{\frac{1}{2}}\) can contain one family
of the charge-conjugated quarks and leptons, which includes a
charge-conjugated neutrino,
\(h \equiv (1,2,2)_{-1}\) contains the MSSM Higgs
doublets \(h_u\) and \(h_d\),
while \(\mathcal{D}\equiv (6,1,1)_{-1}\) contains two
colour-triplet weak-singlet particles, and \(S \equiv (1,1,1)_2\) is a
MSSM singlet.
Including three families contained in three $\mathbf{27}_i$
reps, then, without further constraints on the theory, the allowed
couplings are contained in the tensor products
\cite{Slansky:1981yr,E$_6$SSM}:
\begin{equation} \label{eq:super}
\mathbf{27}_i \mathbf{27}_j
\mathbf{27}_k\rightarrow
F_i F^c_j h_k + F_i F_j \mathcal{D}_k + F^c_i F^c_j \mathcal{D}_k + S_i h_j h_k + S_i \mathcal{D}_j \mathcal{D}_k
\end{equation}
where \(i,j,k = 1 \ldots 3\) are family indices.
However not all these terms are desirable since the presence
of extra Higgs doublets can give rise to flavour changing neutral
currents (FCNCs) and the presence of light colour triplets can
induce proton decay. Therefore extra symmetries are required
to control the couplings, a suitable choice being
the R-symmetry and the discrete $Z^H_2$ symmetry displayed in
Table 4, which reduces the allowed couplings to those
shown in Table 5, where we also display the lowest order
non-renormalizable terms. We now discuss the physics of the allowed and
suppressed terms.

\subsection{Suppressed Flavour Changing Neutral Currents}

We take the \(F_i F^c_j h_3\) superpotential terms to contain the MSSM
Yukawa couplings since we assume that the third generation \(h_3\)
gives the MSSM Higgs doublets.  The other \(h_{\alpha}\) states
are taken to not get VEVs and will cause FCNCs unless the
superpotential term \(F_i F^c_j h_{\alpha}\), where \(\alpha = 1,2\),
is forbidden or highly suppressed by some new symmetry
\cite{E$_6$SSM}.  Here we forbid these terms using a \(Z^H_2\)
discrete symmetry that respects the \(G_{4221}\) symmetry but not the
Planck-scale \(E_6\) symmetry since the latter is assumed to be broken
by quantum gravity.  Under this \(Z^H_2\) symmetry
the `matter particles' \(F_i\) and \(F^c_i\) and `non-Higgs' particles
\(h_{\alpha}\) are taken to have \(Z^H_2 = -1\) and the MSSM Higgs
doublets from \(h_3\) are assumed to have \(Z^H_2 = +1\).  The FCNC
inducing terms \(F_i F^c_j h_{\alpha}\) are therefore forbidden by the
\(Z^H_2\) symmetry and the MSSM superpotential generating terms \(F_i
F^c_j h_3\) are allowed.\footnote{Forbidding or highly suppressing
these terms could explain why only \(h_3\) gets a VEV since then the
other \(h_i\) states won't directly couple to the top Yukawa
coupling.}  

However, we show later that, although the \(F_i F^c_j
h_{\alpha}\) terms are forbidden at the renormalizable level by
\(Z^H_2\), they are still generated from non-renormalizable terms,
which are heavily suppressed so that the induced FCNCs are not
significant.  The \(Z^H_2\) symmetry used here forbids the FCNCs in
the same way that the \(Z^H_2\) symmetry of the E$_6$SSM forbids
the FCNCs from the \(h_{\alpha}\) `non-Higgs' particles in that model
\cite{E$_6$SSM}.

\begin{table}
\begin{center}
 \begin{tabular}{ | c | c |} 
  \hline
  \textbf{Allowed couplings} & \textbf{Physics}
    \\ \hline \(F_i F^c_j h_3\) & MSSM superpotential
    \\ \hline
\(S_3 h_3 h_3\) &  Effective MSSM \(\mu\)-term
\\ \hline
\(S_3 h_{\alpha} h_{\beta}\) & \(h_{\alpha}\) mass
\\ \hline
\(S_3 \mathcal{D}_i \mathcal{D}_j\) & \(\mathcal{D}_i\) mass
\\ \hline
   \(S_{\alpha} h_{\beta} h_3\) & \(S_{\alpha}\) mass
\\ \hline
\(\frac{1}{M_p} \Sigma (F_i F_j \mathcal{D}_k + F^c_i F^c_j
  \mathcal{D}_k) \) & Allows \(\mathcal{D}_i\) and proton decay
\\ \hline
\(\frac{1}{M_P} \Sigma F_i F^c_j h_{\alpha} \)  & Heavily suppressed FCNCs
\\ \hline
\(\frac{1}{M_P} \Sigma S_{\alpha} \mathcal{D}_i \mathcal{D}_j \)  & Harmless
\\ \hline
\(\frac{1}{M_P} \Sigma S_{\alpha} h_{\beta} h_{\gamma} \)  & Harmless
\\ \hline
\(\frac{1}{M_P} \Sigma S_{\alpha} h_{3} h_{3} \)  & Harmless
\\ \hline
\(\frac{1}{M_p} F^c_i F^c_j H_R H_R\) & \(\nu^c\) mass
\\ \hline
 \(\frac{1}{M_p} F_i F_j \overline{H}_L \overline{H}_L\) & Harmless
\\ \hline
\(M (H_R \overline{H}_R + H_L \overline{H}_L)\) & \(\mathbf{16}_H + \overline{\mathbf{16}}_H\) mass
\\ \hline
\end{tabular}
\end{center}
\caption{\footnotesize This table lists the \(G_{4221}\)
superpotential terms that are obtained from all the renormalizable and
first-order non-renormalizable \(E_6\) tensor products of
\(\mathbf{27}_i\), \((\mathbf{16}_H)_{\frac{1}{2}}+(\overline{\mathbf{16}}_{H})_{-\frac{1}{2}}\) (from
a \(\mathbf{27} + \overline{\mathbf{27}}\)), \(M\) and \(\Sigma\) that
are allowed by the \(Z^H_2\) and \(U(1)_R\) symmetries of the
ME$_6$SSM, as discussed in section 5 and table 4.  The indices \(i,j,k
= 1 \dots 3\) and \(\alpha, \beta, \gamma = 1,2\) are family indices.}
\end{table}


\subsection{The $\mu$-Term and Exotic Mass Terms}

As with \(h_i\), we assume that only the third generation of the
\(S_i\) states gets a vacuum expectation value so that the \(S_3 h_3
h_3\) term, from the \(G_{4221}\) superpotential term \(S_i h_j h_k\),
will generate an effective MSSM \(\mu\)-term.  For this term to be
allowed by the \(Z^H_2\) symmetry, the \(S_3\) particles must have
\(Z^H_2 = +1\).  

This \(S_3\) particle is also used to give mass to
the `non-Higgs' particles \(h_{\alpha}\) and colour-triplet particles
\(\mathcal{D}_i\) via the terms \(S_3 h_{\alpha} h_{\beta}\) and \(S_3
\mathcal{D}_i \mathcal{D}_j\) respectively where \(\beta = 1,2\).  For
general \(U(1)'\) models, the \(S_3 \mathcal{D}_i \mathcal{D}_j\)
superpotential term has been shown to induce a VEV for the singlet
\(S_3\) so that it can generate an effective \(\mu\)-term
\cite{Polonsky:1999qd,Dine:1993yw}.  The Yukawa coupling constant for
the \(S_3 \mathcal{D}_i \mathcal{D}_j\) term will, in general,
contribute to the renormalization group evolution of the soft singlet
mass \(m^2_S\) causing it to run negative in the scalar potential.
The VEV of \(S_3\) then carries information about soft supersymmetry
breaking from the parameter \(m^2_S\).  Therefore, the effective
\(\mu\) parameter is now correlated in some way to the SUSY breaking
mechanism and its observed correlation with the soft Higgs mass terms
can be understood \cite{Ibanez:2007pf,Polonsky:1999qd}.  That is, the
\(\mu\) problem of the MSSM should not exist in this model.  We show
below that the \(S_{\alpha} \mathcal{D}_i \mathcal{D}_j\) and
\(S_{\alpha} h_{\beta} h_{\gamma}\) (where \(\gamma = 1,2\))
superpotential terms are forbidden at tee-level so that \(S_{\alpha}\)
should not acquire VEVs. These \(S_{\alpha}\) particles will instead get
mass from the \(S_{\alpha} h_{\beta} h_3\) superpotential terms where
\(S_{\alpha}\) has \(Z^H_2 = -1\).

\subsection{Exotic Decay and Suppressed Proton Decay}

The remaining \(G_{4221}\) superpotential terms to be discussed from
Eq.\ref{eq:super} are \(F_i F_j \mathcal{D}_k\) and \(F^c_i F^c_j
\mathcal{D}_k\).  These will cause rapid proton decay in this model
unless they are highly suppressed or forbidden by some symmetry
\cite{Dimopoulos:1981dw,E$_6$SSM}.  The Standard Model representations of
these superpotential terms are often found to some degree in other
GUTs and the rapid proton decay problems are often solved using some
doublet-triplet splitting mechanism that gives large (above the GUT
scale) mass to the analogue of the \(\mathcal{D}_i\) (triplet)
particles, but EW mass to the Higgs doublets.  However, in our model
we do not give a large mass to the \(\mathcal{D}_i\) particles because
gauge anomalies would then exist, due to the \(U(1)_X\) group, and
Planck scale unification would be lost.  Also, as discussed above, the
\(\mathcal{D}_i\) particles can be used to help induce a VEV for the
\(S_3\) particle, around the EW scale, if they contribute to the low
energy theory. We must therefore highly suppress the \(F_i F_j
\mathcal{D}_k\) and \(F^c_i F^c_j \mathcal{D}_k\) superpotential terms
using a small Yukawa coupling constant rather than using the general
GUT method of creating large \(\mathcal{D}_i\) masses.  

Note that these superpotential terms must be suppressed rather than forbidden
since the \(\mathcal{D}_i\) particles would become stable, strongly
interacting particles with TeV scale masses.  Such particles cannot exist in nature and in fact could potentially cause problems for
nucleosynthesis even if they are unstable with a lifetime greater than just
\(0.1\)s \cite{Wolfram:1978gp}.  Therefore, the \(F_i F_j \mathcal{D}_k\) and \(F^c_i F^c_j
\mathcal{D}_k\) terms should not be suppressed by too small a Yukawa
coupling constant for the lifetime of \(\mathcal{D}_i\) to exceed
\(0.1\)s, or too large a Yukawa coupling constant for the proton's
lifetime to be smaller than the present experimental limits.

To overcome these problems
we use the same method that is used in the model
presented in \cite{Howl:2007hq}.  That is, we forbid the \(F_i F_j
\mathcal{D}_k\) and \(F^c_i F^c_j \mathcal{D}_k\) superpotential terms
at the tree-level but generate them from the non-renormalizable terms
\(\Sigma F_i F_j \mathcal{D}_k\) and \(\Sigma F^c_i F^c_j
\mathcal{D}_k\), where \(\Sigma\) is an \(E_6\) singlet which is
assumed to get a VEV at some high energy scale, by taking both both \(\Sigma\) and \(\mathcal{D}_i\) to have \(Z^H_2 = -1\).  These
non-renormalizable superpotential terms are expected to survive from
the Planck scale and so will likely be suppressed by a factor of
\(1 / M_p\).  We can therefore control the degree of suppression
of the \(F_i F_j \mathcal{D}_k\) and \(F^c_i F^c_j \mathcal{D}_k\)
terms by choosing the energy scale at which \(\Sigma\) gets a VEV.  Below we estimate the energy scales at which \(\Sigma\) can get a VEV so that the proton's lifetime is above the
experimental limits and the \(\mathcal{D}_i\) particles have a
lifetime less than 0.1s.


The superpotential terms \(\lambda_{ijk} F_i F_j \mathcal{D}_k\) and
\(\lambda_{ijk} F^c_i F^c_j \mathcal{D}_k\) cause proton decay through
the decay channels \(p \rightarrow K^{+} \overline{\nu}\) via \(d =
5\) operators (through the \(S_3 \mathcal{D}_i \mathcal{D}_j\) term
which is responsible for the triplet mass $m_D$) and \(p \rightarrow
\pi^0 e^{+}\) via \(d = 6\) operators with matrix elements
proportional to \(\lambda^2 / m_D m_{SUSY}\) and \(\lambda^2 / m^2_D\)
respectively \cite{Nath:2006ut,Wiesenfeldt:2004qa}, where \(m_{SUSY}\)
is the mass scale for the Standard Model's superpartners.  The present
experimental limits on the proton's lifetime for the \(p \rightarrow
K^{+} \overline{\nu}\) and \(p \rightarrow \pi^0 e^{+}\) decay
channels are \(1.6 \times 10^{33}\) years and \(5.0 \times 10^{33}\)
years respectively \cite{Yao:2006px}.  The mass \(m_D\) of the
\(\mathcal{D}\) particles required to suppress the \(p \rightarrow
K^{+} \overline{\nu}\) and \(p \rightarrow \pi^0 e^{+}\) matrix
elements enough for proton decay to not have been observed is given in
various papers (see for example \cite{Raby,Murayama:2001ur}) where no
fine tuning of the Yukawa coupling \(\lambda\) is used.  These
calculations assume that a doublet-triplet splitting mechanism can be
implemented to give large GUT scale masses to the triplet
\(\mathcal{D}\) particles but EW scale masses to the Standard Model
Higgs doublets.  Here we instead assume that the mass of
\(\mathcal{D}\) is not very different from the EW scale
(e.g. \(m_D=1.5\) TeV) and that the Yukawa coupling \(\lambda\) is
very small compared to the Yukawa couplings of the Standard Model.  We
make a rough order of magnitude estimate for the value of the Yukawa
coupling \(\lambda\) required for unobservable proton decay through
the \(d = 5\) and \(d = 6\) channels, for \(m_D=1.5\) TeV, by scaling
the results obtained from \cite{Raby,Murayama:2001ur} and using the
fact that the matrix elements for \(p \rightarrow K^{+}
\overline{\nu}\) and \(p \rightarrow \pi^0 e^{+}\) are proportional to
\(\lambda^2 / m_D m_{SUSY} \) and \(\lambda^2 / m^2_D\) respectively.
In the case of triplets that are much heavier than the doublets, it is the \(d=5\) channel that
sets the higher limit on the mass of the triplets and this is usually
higher than the GUT scale \cite{Raby,Nath:2006ut}.  For \(m_D = 1.5\)
TeV, however, the \(d=5\) and \(d=6\) decay channels have similar
decay rates since \(M_{SUSY}\) is expected to be close to the TeV
scale and the decay rates depend on the square of the matrix elements.
In what follows
we shall choose to scale the results based on the \(d=6\) operator,
since these turn out to give slightly stronger limits on $\lambda$.

According to \cite{Raby} the triplets
\(\mathcal{D}\) must have mass larger than \(10^{10-11}\) GeV for the
lifetime of the proton to be greater than \(5.0 \times 10^{33}\) years
for the non-SUSY \(d=6\) channel \(p \rightarrow K^{+}
\overline{\nu}\) and for no fine-tuning of the Yukawa coupling
constant of the \(F_i F_j \mathcal{D}_k\) and \( F^c_i F^c_j
\mathcal{D}_k\) superpotential terms.  Assuming that the Yukawa
coupling used in \cite{Raby} is of order unity, since it is not
specified, we use the square of the matrix element of \(p \rightarrow
K^{+} \overline{\nu}\) to estimate that the Yukawa coupling
\(\lambda\) must be less than the following when the \(\mathcal{D}\)
triplets have mass equal to \(m_D = 1.5\) TeV:

\[~~~\lambda^4  \lesssim \frac{1}{5.0 \times 10^{33} ~yrs} \times
\bigg(\frac{1.5~\mathrm{TeV}}{10^{11-12}~\mathrm{GeV}} \bigg)^4
\ \ \ \Rightarrow \lambda  \lesssim 10^{-8}.\]

As mentioned above, we forbid the \(F_i F_j
\mathcal{D}_k\) and \(F^c_i F^c_j \mathcal{D}_k\) superpotential terms by the \(Z^H_2\) symmetry but
effectively generate them from the non-renormalizable terms
\(\frac{1}{M_p} \Sigma F_i F_j
\mathcal{D}_k\) and \(\frac{1}{M_p} \Sigma F^c_i F^c_j \mathcal{D}_k\) when \(\Sigma\) gets a
VEV.  To generate an effective Yukawa coupling smaller than \(10^{-8}\) to avoid
experimentally observable proton decay, \(\Sigma\) must get a VEV less than \(10^{11}\) GeV.

The effective superpotential terms \(F_i F_j
\mathcal{D}_k\) and \(F^c_i F^c_j \mathcal{D}_k\), generated from \(\frac{1}{M_p} \Sigma F_i F_j
\mathcal{D}_k\) and \(\frac{1}{M_p} \Sigma F^c_i F^c_j \mathcal{D}_k\), are the only source for the \(\mathcal{D}_i\)
particles to decay. Assuming that \(m_{\widetilde{t}} < m_D\), where \(m_{\widetilde{t}}\) is the mass of the heaviest stop, the \(D^c\) Standard Model representation of the \(G_{4221}\) \(\mathcal{D}\) particle will predominantly decay through the channel \(D^c \rightarrow \widetilde{t} + b\) \cite{E$_6$SSM}.  Using the standard 2-body decay kinematic formula \cite{Yao:2006px} we estimate that the decay rate for \(D^c \rightarrow \widetilde{t} + b\), under the assumption that \(m_b \ll m_{\widetilde{t}}\), is:

\[d \Gamma \sim \frac{1}{32 \pi^2} | \mathcal{M} |^2 \frac{m^2_D - m^2_{\widetilde{t}}}{2 m^3_D} d \Omega\]

At tree-level, a rough order of magnitude estimate of the matrix \(\mathcal{M}\) for the \(D^c \rightarrow \widetilde{t} + b\) decay channel gives:

\[ | \mathcal{M} |^2 \sim 2 ( m^2_D - m^2_{\widetilde{t}}) \lambda^2\]

Taking the mass of the stop to be around the TeV scale, we estimate
that the \(F_i F_j
\mathcal{D}_k\) and \(F^c_i F^c_j \mathcal{D}_k\) operators
must be multiplied by an effective Yukawa coupling \(\lambda\) that is greater than \(10^{-13}\) for the \(\mathcal{D}_i\) particles to have a lifetime less than \(0.1\)s.

If the stop is assumed to be heavier than \(D^c\) (e.g. \(m_{\widetilde{t}} = 2\) TeV, \(m_D = 1.5\) TeV) then the predominant decay channel will most likely be \(D^c \rightarrow b + c + \chi_0\) where \(\chi_0\) is a neutralino \cite{E$_6$SSM}.  Under the assumption that \(m_c \ll m_b \ll m_{\chi_0}\), a tree-level order of magnitude estimate for the decay rate of this channel can be shown to require an effective Yukawa coupling \(\lambda\) about an order of magnitude larger than when the stop is lighter than \(\mathcal{D}\).  We therefore require that \(\lambda \gtrsim 10^{-12}\) for the \(\mathcal{D}_i\) particles to have a lifetime less than \(0.1\)s.

The superpotential terms \(\lambda_{ijk} F_i F_j
\mathcal{D}_k\) and \(\lambda_{ijk} F^c_i F^c_j \mathcal{D}_k\) are effectively generated from the Planck-suppressed operators \(\frac{1}{M_p} \Sigma F_i F_j
\mathcal{D}_k\) and \(\frac{1}{M_p} \Sigma F^c_i F^c_j \mathcal{D}_k\) and so the Yukawa coupling \(\lambda\) is given by \(< \Sigma > / M_p\) where \(< \Sigma >\) is the VEV of the \(\Sigma\) particle.
To avoid cosmological difficulties from the \(\mathcal{D}_i\) particles, \(\Sigma\) must therefore get a
VEV greater than about \(10^7\) GeV.  Therefore, to avoid experimentally observable proton decay and cosmological issues with the \(\mathcal{D}_i\) particles, we require that the \(E_6\) singlet
\(\Sigma\) should get a VEV between \(10^{7-11}\) GeV.


\subsection{R-Symmetry and R-Parity}

To ensure that the LSP is stable in this model, so that it is a
candidate for dark matter, we derive R-parity from the \(U(1)_R\) symmetry
\cite{Barbier:2004ez}, which commutes with the \(G_{4221}\) symmetry
but not the \(E_6\) symmetry because the latter may not be respected
by low energy symmetries as it is assumed to be broken by quantum
gravity effects.
To allow the \(G_{4221}\) superpotential terms, which respect the \(Z^H_2\) discrete symmetry, and to derive a
generalization of the MSSM R-parity, the \(G_{4221}\) supermultiplets
of the three \(\mathbf{27}\) \(E_6\) have the following \(U(1)_R\)
R-charge assignments: \(F_i\) and \(F^c_i\) have \(R = +1\); \(h_3\),
\(h_{\alpha}\), \(\mathcal{D}_i\) and \(\Sigma\) have \(R = 0\); and
\(S_3\) and \(S_{\alpha}\) have \(R = +2\) (see table 4).  The
$16_H$ state also has $R=+2$ so that when it gets a VEV the \(U(1)_R\) is broken to a \(Z_2\) discrete symmetry, which we call
\(Z^R_2\).  Under this $Z^R_2$ symmetry the scalar components of \(F_i\), \(F^c_i\) and the
fermionic components of \(h_3\) (the MSSM sparticles) all have \(Z^R_2
=-1\) while the fermionic components of \(F_i\) and \(F^c_i\) and the
scalar components of \(h_3\) (the MSSM particles) all have \(Z^R_2 =
+1\).  The \(Z^R_2\) symmetry is therefore equivalent to the R-parity
of the MSSM for the \(F_i\), \(F^c_i\) and \(h_3\) supermultiplets.
The \(h_{\alpha}\), \(\mathcal{D}_i\), \(S_i\) and \(\Sigma\)
supermultiplets are not in the MSSM.  All the scalar components of
these `new' supermultiplets can be shown to have \(Z^R_2 = +1\) while
all the fermionic components have \(Z^R_2 = -1\).  Therefore \(F_i\)
and \(F^c_i\) are the only supermultiplets in the theory which have
\(Z^R_2 = +1\) for their fermionic components and \(Z^R_2 = -1\) for
their scalar components.  This \(Z^R_2\) symmetry stops the `non-MSSM'
particles from allowing the MSSM LSP to decay as well as operating as
the R-parity of the MSSM.  
The introduction of the $Z^R_2$ symmetry therefore 
ensures a stable dark matter candidate, the MSSM LSP.

Note that the $Z^H_2$ symmetry in table 4 is equivalent to an MSSM
matter-parity. Therefore if it was left unbroken then it
would also prevent the MSSM LSP from decaying.  However, as discussed
in section 5.3, the $Z_2^H$ symmetry is broken by the $E_6$ singlet
$\Sigma$ at around $10^{7-11}$ GeV generating the effective operators
\(F_i F_j \mathcal{D}_k\), \(F^c_i F^c_j \mathcal{D}_k\) and \(F_i
F^c_j h_{\alpha}\) that disrespect \(Z^H_2\), and enabling the MSSM
LSP to decay.  Hence the $Z^R_2$ symmetry must be introduced in
addition to the $Z^H_2$ symmetry so that the MSSM LSP is stable.

\subsection{\(\mathbf{16}_H + \overline{\mathbf{16}}_H\) Mass}

In addition to the three \(\mathbf{27}\) \(E_6\) multiplets, which are
present at low energies, the Higgs states from \((\mathbf{16}_H)_{\frac{1}{2}}+(\overline{\mathbf{16}}_{H})_{-\frac{1}{2}}\) or \(\mathbf{27}_H +
\overline{\mathbf{27}}_H\) must be given a mass at the conventional GUT
scale so that the \(G_{4221}\) symmetry can be broken and the gauge
coupling constants can unify at the Planck scale.  Here we only consider
a model that contains the \((\mathbf{16}_H)_{\frac{1}{2}}+(\overline{\mathbf{16}}_{H})_{-\frac{1}{2}}\)
above the GUT scale and briefly mention what changes should be made to
such a model if \(\mathbf{27}_H + \overline{\mathbf{27}}_H\) are used
to break the \(G_{4221}\) symmetry instead.  

To give the required GUT scale masses to the
\((\mathbf{16}_H)_{\frac{1}{2}}+(\overline{\mathbf{16}}_{H})_{-\frac{1}{2}}\)
multiplets, we introduce an \(E_6\) singlet \(M\) that is assumed to
get a VEV at this particular energy scale and which couples to the
\(G_{4221}\) representations of the \((\mathbf{16}_H)_{\frac{1}{2}}\)
and \((\overline{\mathbf{16}}_{H})_{-\frac{1}{2}}\) multiplets via the
superpotential term \(M (\mathbf{16}_H)_{\frac{1}{2}}
(\overline{\mathbf{16}}_H)_{-\frac{1}{2}}\).  We take the
\((\mathbf{16}_H)_{\frac{1}{2}}\) supermultiplet to have an R-charge
of \(+2\) so that certain phenomenologically problematic operators are
forbidden. Then, to
allow the \(M (\mathbf{16}_H)_{\frac{1}{2}}
(\overline{\mathbf{16}}_H)_{-\frac{1}{2}}\) term, the \(M\) and
\((\overline{\mathbf{16}}_H)_{-\frac{1}{2}}\) supermultiplets are both
given an R-charge of \(0\).\footnote{Note that the bilinear term \(
(\mathbf{16}_H)_{\frac{1}{2}}
(\overline{\mathbf{16}}_H)_{-\frac{1}{2}}\) is also allowed by the
symmetries of the model.  We assume that the dimensional coupling
constant for this term is less than or equal to \(M_{GUT}\).} If
\(\mathbf{27}_H + \overline{\mathbf{27}}_H\) is included between the
GUT and Planck scales, rather than
\((\mathbf{16}_H)_{\frac{1}{2}}+(\overline{\mathbf{16}}_{H})_{-\frac{1}{2}}\),
then another \(Z_2\) must be applied to the \(\mathbf{27}_H\) and
\(\overline{\mathbf{27}}_H\) multiplets to forbid certain
phenomenologically problematic superpotential terms.

\subsection{Neutrino Mass}

These R-charge assignments forbid
phenomenologically-problematic terms and allow the charge-conjugated
neutrinos, from \(F^c_i\), to obtain a large Majorana mass
\(\mathcal{O} (M^2_{GUT} / M_p)\)
from a \(\frac{1}{M_p}
F^c_i F^c_j (\overline{\mathbf{16}}_H)_{-\frac{1}{2}} (\overline{\mathbf{16}}_H)_{-\frac{1}{2}} \equiv
\frac{1}{M_p} F^c_i F^c_j H_R H_R\) superpotential term.  This term
will create a conventional see-saw mechanism for the left-handed
neutrinos when \(h_3\) gets a VEV in the superpotential term \(F_i
F^c_j h_3\).  This and \(\frac{1}{M_p} F_i F_j \overline{H}_L
\overline{H}_L\), which is phenomenologically harmless, are the only
superpotential terms that contain interactions between the three
\(\mathbf{27}\) \(E_6\) multiplets and the \((\mathbf{16}_H)_{\frac{1}{2}}+(\overline{\mathbf{16}}_{H})_{-\frac{1}{2}}\) and \(M\) multiplets.



\section{Summary and Conclusion}
We have proposed a Minimal $E_6$ Supersymmetric Standard Model (ME$_6$SSM)
based on three low energy reducible $\mathbf{27}$ representations of the
Standard Model gauge group which has many attractive features
compared to the MSSM. In particular it provides a solution to the
$\mu$ problem and doublet-triplet splitting problem,
without re-introducing either of these problems.
Above the conventional GUT scale the ME$_6$SSM is
embedded into a left-right symmetric Supersymmetric Pati-Salam model,
which allows complete gauge unification at the Planck scale,
subject to gravitational uncertainties. Although we have not studied
it here, it is also clear that fine-tuning in such models
will be significantly reduced compared to the MSSM
due to the enhanced Higgs mass.

At low energies there is an additional $U(1)_X$ gauge group,
consisting of a novel and non-trivial
linear combination of one Abelian and two non-Abelian Pati-Salam
generators. The $U(1)_X$
is broken at the TeV scale by the same singlet
that also generates the effective \(\mu\) term, resulting in a
new low energy \(Z'\) gauge boson.
We compared the \(Z'\) of the ME$_6$SSM (produced via the
Pati-Salam breaking chain of $E_6$, where $E_6$ is broken
at the Planck scale)
to the  \(Z'\) of the E$_6$SSM (from the $SU(5)$ breaking chain
of $E_6$, where $E_6$ is broken
at the GUT scale) and discussed how they can be distinguished by their
different couplings. The possible discovery of such \(Z'\)
gauge bosons is straightforward at the LHC and the different
couplings should enable the two models to be resolved experimentally.
In particular, the most significant difference between the vector and
axial couplings of the \(Z'\) of
the E$_6$SSM and ME$_6$SSM is in the vector
coupling of the charges leptons, which is twice as large in the
ME$_6$SSM as in the E$_6$SSM. 

We emphasise that
the presence of additional threshold
corrections at the Planck scale will not change the Pati-Salam
breaking scale or the values of the
Standard Model gauge couplings at this scale to one loop order.
However, since these quantities are determined by running up the
couplings from low energies, there will be some sensitivity
to TeV scale threshold corrections. Since the vector and axial
vector couplings of the $Z'$ are determined from the values 
of the Standard Model gauge couplings at the Pati-Salam
breaking scale, there will therefore be little
sensitivity to Planck scale threshold corrections on the 
determined vector and axial vector couplings of the $Z'$.

We have introduced an R-symmetry and discrete \(Z^H_2\) symmetry that addresses
the potential major phenomenological
problems such as flavour changing neutral currents and proton decay,
which would otherwise be introduced to the theory by colour triplet fermions
and extra non-Higgs doublets from the three copies of the \(\mathbf{27}\)
multiplet. In the ME$_6$SSM, right-handed Majorana masses
of the correct order of magnitude
naturally arise from the Higgs mechanism that breaks the intermediate
Pati-Salam and \(U(1)_{\psi}\) symmetry to the standard model and
\(U(1)_X\) gauge group, leading to a conventional see-saw mechanism.
It should be possible to embed the model presented here into
a realistic flavour model describing all quark and lepton masses,
leading to predictions for the exotic colour triplet non-Higgs
fermion masses, which will be the subject of a future study.

In conclusion, the ME$_6$SSM has clear advantages over both the
MSSM and NMSSM, and even the E$_6$SSM, which make it a serious
candidate SUSY Standard Model.
It also has a certain elegance in the way that the low energy theory
contains only complete reducible $\mathbf{27}$ representations that also allow for anomaly cancellation of the gauged
$U(1)_X$, which we find quite appealing. We have shown that the
potentially dangerous couplings of the exotic particles can readily be tamed
by simple symmetries, leading to exciting predictions
at the LHC of exotic colour triplet fermions and a new \(Z'\) with
distinctive couplings. The discovery and study of such new
particles could provide a glimpse into the physics
of unification at the Planck scale.

\subsection*{Acknowledgments}
RJH acknowledges a PPARC Studentship.
SFK acknowledges partial support from the following grants:
PPARC Rolling Grant PPA/G/S/2003/00096;
EU Network MRTN-CT-2004-503369;
EU ILIAS RII3-CT-2004-506222;
NATO grant PST.CLG.980066.

\section*{Appendix: Symmetry Breaking \(G_{4221}\) to  \(G_{3211}\)}

Since the \(U(1)_X\)
group does not appear to have been considered in the literature, we
illustrate in detail how the \(H_R\) and \(\overline{H}_R\) Higgs
bosons can generate it and also break the \(G_{4221} = SU(4) \otimes SU(2)_L \otimes SU(2)_R \otimes U(1)_{\psi}\) symmetry to the \(G_{3211} = SU(3)_c \otimes
SU(2)_L \otimes U(1)_Y \otimes U(1)_X\) symmetry.

We start with the
covariant derivative of the \(G_{4221}\) symmetry, which can be
written as:

\begin{equation} \label{eq:covariant derivative}
D_{\mu} = \partial_{\mu}  + ig_4 T_4^m A^m_{4 \mu} + ig_{2L} T^s_L
A^s_{L \mu} + ig_{2R} T^r_R
A^r_{R \mu} +  \frac{1}{\sqrt{6}} i g_{\psi} T_{\psi} A_{\psi \mu}
\end{equation}

where \(m = 1 \ldots 15\) and \(r,s = 1 \ldots 3\);  \(A^{m}_{4
\mu}\), \(A^s_{L \mu}\), \(A^r_{R \mu}\) and \(A_{\psi \mu}\) are the \(SU(4)_c\), \(SU(2)_L\),
\(SU(2)_R\) and \(U(1)_{\psi}\) quantum fields respectively;
\(g_4\), \(g_{2L}\), \(g_{2R}\) and \(g_{\psi}\) denote the universal gauge coupling
constants of the respective fields and \(T^m_4\), \(T^s_L\), \(T^r_R\) and
\(T_{\psi}\) represent their generators.  All of the
\(T^m_4\), \(T^r_R\), \(T^s_L\) and \(T_{\psi}\) generators are derived from components of the \(E_6\) generators \(G^a\), which we choose to \(E_6\) normalize, for the fundamental representation \(\mathbf{27}\), by:

\begin{equation} \label{eq:norm}
 \mathrm{Tr}(G^a~ G^b) = 3 \delta^{ab}
\end{equation}

where \(a,b = 1 \ldots 78\).

Then, with this normalization, the Pati-Salam generators \(T^m_4\), \(T^r_R\) and \(T^s_L\) are normalized for the fundamental representations of \(SU(4)\), \(SU(2)_R\) and \(SU(2)_L\) respectively, by:\footnote{These normalizations are necessary for the standard model
generators \(T_{SM}\) of \(SU(3)_c\) and \(SU(2)_L\) to be
normalized in the conventional way: \(Tr(T^d_{c} T^e_{c}) =
\frac{1}{2} \delta^{de}\) and \(Tr(T^r_{L} T^s_{L}) = \frac{1}{2} \delta^{rs}\) for the fundamental representations, where \(T_{c}\) and \(T_{L}\) are the generators for the \(SU(3)_c\) and \(SU(2)_L\) groups respectively and \(d,e = 1 \ldots 8\). }

\[Tr(T^m_4 ~T^n_4) = \frac{1}{2} \delta^{mn}, \]
\[Tr(T^r_R ~T^s_R) = Tr(T^r_L ~T^s_L) = \frac{1}{2} \delta^{rs}\]

where \(m,n = 1 \ldots 15\).

The \(U(1)_{\psi}\) charge \(\frac{1}{\sqrt{6}} T_{\psi}\) is a diagonal \(E_6\) generator, which we choose to be the 78th generator \(G^{78} = \frac{1}{\sqrt{6}} T_{\psi} \), and is therefore normalized by Eq.\ref{eq:norm} to give:

\begin{equation} \label{eq:normTpsi}
\frac{1}{6} \sum_{\mathbf{27}} T^2_{\psi} = 3
\end{equation}

where the sum is over all the \(G_{4221}\) representations that make
up the fundamental \(\mathbf{27}\) multiplet of \(E_6\).

To break \(G_{4221}\) to \(SU(3)_c \otimes SU(2)_L \otimes U(1)_Y
\otimes U(1)_X\) we use the Higgs bosons \(H_R\) and \(\overline{H}_R\) that transform as
\((4,1,2)_{-\frac{1}{2}}\) and \((\overline{4},1,\overline{2})_{\frac{1}{2}}\) under \(G_{4221}\) respectively.
These are the smallest \(G_{4221}\)
multiplets that can be used to break the Pati-Salam symmetry directly
to the standard model gauge group.  When \(H_R\) and \(\overline{H}_R\) develop VEVs in the \(\nu_R\) and
\(\nu^c\) components respectively, they will break \(SU(4)_c
\rightarrow SU(3)_c\) \cite{Pati:1974yy} and mix the field associated with the remaining
\(SU(4)_c\) diagonal generator, \(A^{15}_4\), with the field
associated with the diagonal generator of \(SU(2)_R\), \(A^3_R\), and
the \(U(1)_{\psi}\) field \(A_{\psi}\).  The rest of the \(SU(4)_c\)
and \(SU(2)_R\) fields are given square mass proportional to
\(v^2\), the sum of the square of the \(H_R\) and
\(\overline{H}_R\) VEVs.

The diagonal generators for the \(A^{15}_4\) and \(A^3_R\) fields are \(T^{15}_4\) and \(T^3_R\).  For the fundamental representations of
\(SU(4)\) and \(SU(2)_R\) respectively \cite{Slansky:1981yr} :

\[T^{15}_4 =
\sqrt{\frac{3}{2}}~diag(\frac{1}{6}, \frac{1}{6}, \frac{1}{6},
-\frac{1}{2}),~~~~T^3_R = diag(\frac{1}{2}, -\frac{1}{2}).\]

The part of the symmetry breaking
\(G_{4221}\) to  \(G_{3211}\)
involving the diagonal generators \(T^{15}_4\), \(T^3_R\) and \(T_{\psi}\) is then equivalent to:

\[U(1)_{T^{15}_4} \otimes U(1)_{T^3_R} \otimes U(1)_{\psi} \rightarrow U(1)_Y \otimes U(1)_X.\]

In the rest of this Appendix we explain this particular symmetry breaking in detail.  Using the \(G_{4221}\) covariant derivative, Eq.\ref{eq:covariant derivative},
the covariant derivative for the \(U(1)_{T^{15}_4} \otimes U(1)_{T^3_R} \otimes U(1)_{\psi}\)
symmetry is:

\[D_{\mu} = \partial_{\mu} + i g_4 T^{15}_4 A^{15}_{4 \mu} + i g_{2R} T^3_R A^3_{R \mu} + \frac{1}{\sqrt{6}} i g_{\psi} T_{\psi} A_{\psi \mu}\]

\begin{equation} \label{eq:G111}
~~~~\equiv \partial_{\mu} + i g_{B-L} T_{B-L} A^{15}_{4 \mu} + i g_{2R} T^3_R A^3_{R \mu} + i g_{N \psi} T_{\psi} A_{\psi \mu}
\end{equation}

where \(g_{B-L} \equiv \sqrt{\frac{3}{2}} g_4\), \(g_{N \psi} \equiv \frac{1}{\sqrt{6}} g_{\psi}\), \(T_{B-L} \equiv \sqrt{\frac{2}{3}} T^{15}_4 = \frac{(B-L)}{2}\) and \(B\) and \(L\) are baryon and lepton number respectively.

In terms of the diagonal generators \(T_{B-L}\), \(T^3_R\) and
\(T_{\psi}\), the \(\nu_R\) component of \(H_R\) and the \(\nu^c\)
component of \(\overline{H}_R\) transform under \(U(1)_{T^{15}_4}
\otimes U(1)_{T^3_R} \otimes U(1)_{\psi}\) as:

\be
\nu_{R}^H=\left(-\frac{1}{2}, \ \frac{1}{2},\
-\frac{1}{2}\right),~~~~~\nu^{c}_{\overline{H}} = \left(\frac{1}{2}, \ -\frac{1}{2},\
\frac{1}{2}\right).
\ee

Therefore, once
\(H_R\) and \(\overline{H}_R\) get their VEVs, the square of the covariant derivative for the \(A^{15}_4\),
\(A^3_R\) and \(A_{\psi}\) fields becomes:

\[\Big| D_{\mu} \nu_{R}^H \Big|^2 =  \frac{1}{4} v^2 \bigg(- \mathbf{g}_{B-L} A^{15}_{4 \mu} +\mathbf{g}_{2R} A^3_{R \mu} - \mathbf{g}_{N\psi} A_{\psi \mu}  \bigg)^2 \]

 where \(\mathbf{g}_{B-L} \), \(\mathbf{g}_{2R}\) and \(\mathbf{g}_{N\psi}
\) are the \(g_{B-L}\), \(g_{2R}\) and \(g_{N \psi}\) gauge coupling constants evaluated at the \(G_{4221}\) symmetry breaking scale.
 The above squared covariant derivative can be written in matrix form as:

\begin{equation} \label{eq:matrix}
  \frac{1}{4} v^2 \left( \begin{array} {ccc} A^3_R & A^{15}_4 & A_{\psi} \end{array} \right) \left ( \begin{array} {ccc} \mathbf{g}^2_{2R} &
  -\mathbf{g}_{2R} ~\mathbf{g}_{B-L} & -\mathbf{g}_{2R} ~\mathbf{g}_{N
    \psi}\\ -\mathbf{g}_{2R} ~\mathbf{g}_{B-L} & \mathbf{g}^2_{B-L} &
  \mathbf{g}_{B-L} ~\mathbf{g}_{N\psi}\\ -\mathbf{g}_{2R}
  ~\mathbf{g}_{N\psi} & \mathbf{g}_{B-L} ~\mathbf{g}_{N\psi} &
  \mathbf{g}^{2}_{N\psi}
\end{array} \right ) \left ( \begin{array} {c} A^3_R \\ A^{15}_4 \\ A_{\psi} \end{array} \right).
\end{equation}

Diagonalizing this matrix equation determines the mass eigenstate fields generated by the mixing
of the \(G_{4221}\) fields \(A^3_R\), \(A^{15}_4\) and \(A_{\psi}\).  The \(3 \times 3\) square mass
mixing matrix has two zero eigenvalues and one non-zero eigenvalue so that two
massless gauge bosons and one massive gauge boson appear to have been created by
the mixing.  The massive gauge boson \(B_H\) is the following mixture of \(G_{4221}\)
fields:

\[B_H = \frac{1}{b} \Big( - \mathbf{g}_{2R} A^3_R + \mathbf{g}_{B-L} A^{15}_4 + \mathbf{g}_{N \psi} A_{\psi} \Big)\]

where \(b^2 \equiv \mathbf{g}^2_{2R} + \mathbf{g}^2_{B-L} + \mathbf{g}^2_{N\psi}\).

This massive field is an unique mass eigenstate field.  However, the degeneracy in
the zero-eigenvalue eigenvectors of the square mass mixing matrix implies that all
orthogonal combinations of any chosen two massless eigenstate fields also describe
two massless eigenstate fields.  All the orthogonal combinations of two massless eigenstate fields are physically distinct and so the symmetry breaking mechanism does not generate two unique massless eigenstate fields.  We therefore require something in addition to this symmetry breaking mechanism that lifts the degeneracy of the zero-eigenvalue eigenvectors and selects two unique massless gauge fields.

We show below that when we include the low-energy VEV of the \(S\) particle
from the third generation of the \(\mathbf{27}\) multiplets, the degeneracy in the zero-eigenvalue eigenvectors is lifted and the two massless gauge fields are uniquely chosen to be the gauge field \(B_Y\) of the Standard
Model hypercharge group and an (effectively massless) gauge field that we call \(B_X\).
The \(B_Y\) and \(B_X\) gauge fields are generated from orthogonal zero-eigenvalued
eigenvectors of the above \(3 \times 3\) square mass mixing matrix and are the following mixture
of \(G_{4221}\) fields:

\[B_Y = \frac{1}{a} \Big( \mathbf{g}_{B-L} A^3_R + \mathbf{g}_{2R} A^{15}_4 \Big),\]

\[B_X = \frac{1}{ab} \Big(  \mathbf{g}_{2R} \mathbf{g}_{N \psi} A^3_R - \mathbf{g}_{B-L} \mathbf{g}_{N \psi} A^{15}_4
 + (\mathbf{g}^2_{2R} + \mathbf{g}^2_{B-L}) A_{\psi} \Big)\]

where \(a^2 \equiv \mathbf{g}^2_{2R} + \mathbf{g}^2_{B-L}\).

In terms of the diagonal generators \(T_{B-L}\), \(T^3_R\) and \(T_{\psi}\), the \(S\)
particle transforms under the \(U(1)_{T^{15}_4} \otimes U(1)_{T^3_R} \otimes U(1)_{\psi}\) symmetry as:

\[
S = \Big(0,~ 0,~
2 \Big).
\]

The \(S\) particle only couples to \(A_{\psi}\) and so its VEV, \(s\), therefore
introduces a perturbation proportional to \(s^2 / v^2\) to the \(33\) component of
the \(3 \times 3\) square mass mixing matrix in Eq.\ref{eq:matrix}. From Section 3, \(v\)
is determined to be of the order \(10^{16}~\mathrm{GeV}\) and we require that
\(s \sim 10^{3}~ \mathrm{GeV}\) for EW symmetry breaking.

Diagonalizing the \(3 \times 3\) square mass mixing matrix with this extremely small perturbation in the
\(33\) component determines the mass eigenstate fields to be the massless hypercharge gauge field \(B_Y\),
and an extremely small mass gauge field and large mass gauge field that can be taken to be the \(B_X\) and \(B_H\) gauge fields, respectively, in the excellent approximation that \(s^2 / v^2 = 0\).\footnote{We have ignored the VEV of the Standard Model Higgs boson in this symmetry breaking.}

It is easy to see why the hypercharge gauge field of the Standard Model is the exact massless gauge field of this symmetry breaking. The hypercharge field is the only massless gauge field generated by the \(H_R\) and \(\overline{H}_R\) VEVs that does not contain
the \(A_{\psi}\) field and therefore the only massless gauge field that \(S\) does not couple to.
If the \(A_{\psi}\) field is removed from the \(G_{4221}\) symmetry then the mixing of the remaining \(G_{4221}\) diagonal generators becomes
equivalent to \(U(1)_{T^{15}_4} \otimes U(1)_{T^3_R} \rightarrow U(1)_Y\) when
\(H_R\) and \(\overline{H}_R\) get VEVs \cite{Howl:2007hq,Pati:1974yy}.

The mass eigenstate fields \(B_Y\), \(B_X\) and \(B_H\) can be written in terms of the \(G_{4221}\) fields \(A^3_R\), \(A^{15}_4\) and \(A_{\psi}\)
in the following matrix form:

\begin{equation} \label{eq:VT}
\left( \begin{array} {c} B_Y \\ B_X \\ B_H \end{array} \right)  = \left ( \begin{array} {ccc} \mathbf{g}_{B-L} / a & \mathbf{g}_{2R} / a  & 0\\
\mathbf{g}_{2R} \mathbf{g}_{N \psi} / ab  & -\mathbf{g}_{B-L} \mathbf{g}_{N \psi} / ab & (\mathbf{g}^2_{2R} + \mathbf{g}^2_{B-L}) / ab \\
 -\mathbf{g}_{2R} / b & \mathbf{g}_{B-L} / b & \mathbf{g}_{N \psi} / b
\end{array} \right ) \left( \begin{array} {c} A^3_R \\ A^{15}_4 \\ A_{\psi} \end{array} \right).
\end{equation}


We choose to parameterize this orthogonal \(3 \times 3\) matrix in terms of rotation and reflection matrices in the following way:

\[\left( \begin{array} {ccc} 1 & 0 & 0
                          \\ 0 & c_{23} & s_{23}
                          \\ 0 & -s_{23} & c_{23} \end{array} \right)
 \left( \begin{array} {ccc} 1 & 0 & 0
                          \\ 0 & -1 & 0
                          \\ 0 & 0 & 1 \end{array} \right)
 \left( \begin{array} {ccc} c_{12} & s_{12} & 0
                          \\ -s_{12} & c_{12} & 0
                          \\ 0 & 0 & 1 \end{array} \right)
=\left( \begin{array} {ccc} c_{12} & s_{12} & 0
                          \\ c_{23} s_{12} & -c_{23} c_{12} & s_{23}
                          \\ -s_{23} s_{12} & s_{23} c_{12} & c_{23} \end{array} \right)
\]

where \(c_{12} = \mathbf{g}_{B-L} / a\), \(s_{12} =  \mathbf{g}_{2R} / {a}\), \(c_{23} = \mathbf{g}_{N \psi} / {b}\) and \(s_{23} = a / b\).  The mixing angles \(\theta_{12}\) and \(\theta_{23}\) are therefore given by \(\tan \theta_{12} =   \mathbf{g}_{2R} / \mathbf{g}_{B-L}\) and \(\tan \theta_{23} = a / \mathbf{g}_{N\psi}\).

Taking the transpose of Eq.\ref{eq:VT}, the \(G_{4221}\) fields \(A^3_R\), \(A^{15}_4\) and \(A_{\psi}\) can be written in terms of the mass eigenstate fields \(B_Y\), \(B_X\) and \(B_H\) as:

\[
\left( \begin{array} {c} A^3_R \\ A^{15}_4 \\ A_{\psi} \end{array} \right) = \left ( \begin{array} {ccc}
c_{12}         & s_{12} c_{23}    &   -s_{12} s_{23} \\
s_{12}         & -c_{12} c_{23}     &  c_{12} s_{23}\\
 0             & s_{23}   &   c_{23}
\end{array} \right ) \left( \begin{array} {c} B_Y \\ B_X \\ B_H \end{array} \right).\]

Putting this matrix equation into the covariant derivative for the  \(U(1)_{T^{15}_4} \otimes U(1)_{T^3_R} \otimes U(1)_{\psi}\) symmetry, Eq.\ref{eq:G111}, determines the covariant derivative for the massless gauge fields \(B_Y\) and \(B_X\) to be:

\[D_{\mu} = \partial_{\mu} + i g_Y Y B_{Y \mu} + i g^0_{X} X B_{X \mu}\]

where:

\[Y = T^3_R + T_{B-L} = T^3_R + (B-L) / 2\]

is the Standard Model hypercharge \cite{Pati:1974yy},

\begin{equation} \label{eq:Z}
X = (T_{\psi} + T^3_R) - c^{2}_{12} Y
\end{equation}

is the non-normalized charge of the \(B_X\) gauge field. \(g_Y\) and \(g^0_{X}\) are the non-normalized universal gauge coupling constants of the \(B_Y\) and \(B_X\) fields respectively and, at the \(G_{4221}\) symmetry breaking scale, are given by Eq.\ref{eq:gY} and Eq.\ref{eq:g0X}:\footnote{Note that Eq.\ref{eq:gY} is the relation that \(g_Y\) must satisfy if the Pati-Salam symmetry, without the \(U(1)_{\psi}\), was broken to the standard model gauge group using a Higgs boson that transforms as \((4,1,2)\) and gets a VEV in the \(\nu_R\) direction \cite{Howl:2007hq}.
}

\begin{equation} \label{eq:gY}
g_Y =  \frac{\mathbf{g}_{2R}
~\mathbf{g}_{B-L} }{a}
\end{equation}

\begin{equation} \label{eq:g0X}
g_X^0 = \frac{a}{b} \mathbf{g}_{N\psi}.
\end{equation}

Eq.\ref{eq:gY} and Eq.\ref{eq:g0X} can be written is terms of \(\alpha_Y = \frac{g^2_Y}{4 \pi}\) and \(\alpha^0_X = \frac{(g^0_X)^2}{4 \pi}\), see Eq.\ref{eq:relation} and Eq.\ref{eq:alphaX} in Section 3.

The charges \(X\) and \(Y\) are not \(E_6\) normalized.  We write the normalized respective charges as \(T_X\) and \(T_Y\) where:

\[T_X = X / N_X,~~~~~T_Y = Y / N_Y\]

and the normalization constants \(N_X\) and \(N_Y\)  are given by:

\[N^2_X = 7 - 2c^2_{12} + \frac{5}{3} c^4_{12}, ~~~~ N^2_Y = \frac{3}{5}\]

Note that the Abelian generator \(T_Y\) is just the conventional GUT normalized hypercharge. \(T_X\) and \(T_Y\) have been \(E_6\) normalized using Eq.\ref{eq:norm} which is equivalent to:

\[\sum_{\mathbf{27}} T^2_Y = \sum_{\mathbf{27}} T^2_X =  3\]

where the sum is over all the \(G_{3211} \equiv SU(3)_c \otimes SU(2)_L \otimes U(1)_Y \otimes U(1)_X\) representations of the fundamental \(\mathbf{27}\) \(E_6\) multiplet and \(U(1)_X\) is the unitary group of the \(B_X\) field.\footnote{We could have defined \(X\) and \(N_X\) differently as long as \(T_X\) is the same. Here we have chosen to define \(X\) so that it can be written in terms of hypercharge \(Y\).}

In terms of the \(E_6\) normalized charges \(T_X\) and \(T_Y\), the covariant derivative for the \(B_X\) and \(B_Y\) gauge fields becomes:

\begin{equation} \label{eq:UYUX}
D_{\mu} = \partial_{\mu} + i g_1 T_Y B_{Y \mu} + i g_X T_X B_{X \mu}
\end{equation}

where \(g_1\) and \(g_{X}\) are the \emph{normalized} universal gauge coupling constants of the \(B_Y\) and \(B_X\) fields respectively.  At the \(G_{4221}\) symmetry breaking scale, the normalized gauge coupling constants \(g_1\) and \(g_X\) are the following combinations of \(G_{4221}\) gauge coupling constants:

\[
g_1  = N_Y \frac{\mathbf{g}_{2R}
~\mathbf{g}_{B-L} }{a},~~~~~~
g_{X} = N_X ~\frac{a}{b} \mathbf{g}_{N\psi}.
\]

From Eq.\ref{eq:Z}, the charge \(T_X\) of the \(U(1)_X\) group depends on the Pati-Salam
gauge coupling constants \(g_{2R}\) and \(g_{B-L}\) evaluated at the
\(G_{4221}\) symmetry breaking scale.  Therefore, under the excellent approximation that \(s^2 / v^2 = 0\), a massless gauge boson exists that couples to
particles with a charge that depends on the values that certain
coupling constants take at some high energy scale.  Although this may be unusual, it does not appear to pose any problems.  Indeed, like
any other quantum charge, \(T_X\) is a dimensionless constant
that is independent of the energy scale at which the interaction
between the particle and the \(A_{X}\) field occurs and, although the
numbers that \(X\)
takes may not be
able to be arranged into fractions like \(Y\),
they are still discrete and sum to zero for a complete \(E_6\)
representation. However, unlike conventional \(U(1)\) charges, \(T_X\) is obviously very model dependent since different \(E_6\) models with an intermediate Pati-Salam symmetry will, in general, contain different values of the gauge coupling constants \(g_{2R}\) and \(g_4\) evaluated at the \(G_{4221}\) symmetry breaking scale.  It is easy to prove that it is a general rule that, if three massless gauge
fields are mixed, then at least two of the resulting mass eigenstate
 fields must have a charge that
depends on the value of the original gauge coupling constants.  Therefore this gauge coupling dependence is not peculiar to the Higgs symmetry breaking mechanism discussed in this Appendix, but to any symmetry breaking mechanism involving three fields.

In this Appendix we have illustrated how the \(G_{4221} \equiv SU(4) \otimes SU(2)_L \otimes SU(2)_R \otimes U(1)_{\psi}\) symmetry can be broken to the symmetry \(G_{3211} \equiv SU(3)_c \otimes SU(2)_L \otimes U(1)_Y \otimes U(1)_X\) when the \(G_{4221}\) multiplets \(H_R\), \(\overline{H}_R\) and \(S\) get vacuum expectation values.  Using the covariant derivatives for the \(G_{4221}\) symmetry, Eq.\ref{eq:covariant derivative}, and the \(U(1)_Y \otimes U(1)_X\) symmetry, Eq.\ref{eq:UYUX}, the covariant derivative for the \(G_{3211}\) symmetry is given by:

\begin{equation} \label{eq:DmuG3211}
D_{\mu} = \partial_{\mu} + ig_{3} T^n_{3c} A^n_{3c \mu} +
ig_{2L} T^s_{L} A^s_{L \mu} + i g_1 T_Y B_{Y \mu} + i g_X T_X B_{X \mu}
\end{equation}

where \(A^n_{3c}\) and \(T^n_{3c}\) are the  \(SU(3)_c\) fields and
generators derived from the \(SU(4)\) symmetry respectively (with \(n = 1 \ldots
8\)) and \(g_{3c}\) is the universal gauge coupling constant of  \(A^n_{3c}\).

We consider this \(G_{3211}\) symmetry as an effective high energy symmetry under the assumption that the low-energy VEVs of the MSSM singlet \(S\) and MSSM Higgs bosons can be neglected at higher energy scales.

\end{document}